\documentclass[a4paper,12pt]{article}
\usepackage[utf8]{in
    putenc}
\usepackage{cancel}
\usepackage{ulem}
\usepackage{amsfonts}
\usepackage{amssymb}
\usepackage{graphicx}
\usepackage{amsmath,bm}
\usepackage{enumerate}
\usepackage{mathtools}
\usepackage{tikz} \usetikzlibrary{calc}
\usepackage[countmax]{subfloat}
\usepackage{xcolor}
\usepackage{float}
\usepackage{color}
\usepackage{here}
\usepackage{cite}
\usepackage{subcaption}
\usepackage{mwe}

\usepackage{mathrsfs}
\usepackage{float,epsfig}
\usepackage{dcolumn}
\usepackage{pgfplots}
\usepackage{graphicx}
\usepackage{bm}
\usepackage{amsmath,amssymb,amsthm}
\usepackage[colorlinks=true,linkcolor=blue,citecolor=red]{hyperref}
\definecolor{lime}{HTML}{A6CE39}
\newcommand{\orcidicon}{%
    \begin{tikzpicture}
    \draw[lime, fill=lime] (0,0)
        circle [radius=0.16]
        node[white] {{\fontfamily{qag}\selectfont \tiny ID}};
    \draw[white, fill=white] (-0.0625,0.095)
        circle [radius=0.007];
    \end{tikzpicture}   \hspace{-2mm}
}

\newcommand\orcidHasan{{\href{https://orcid.org/0000-0001-7408-0910}{\orcidicon}}}
\newcommand\orcidKarima{{\href{https://orcid.org/0000-0001-5419-8516}{\orcidicon}}}
\newcommand\orcidFaical{{\href{https://orcid.org/0000-0002-2977-0821}{\orcidicon}}}

\title{\bf On some phase equilibrium features of charged black
holes in flat spacetime via Rényi statistics}

\author{
F. Barzi\orcidFaical\!\!$^{1,3}$\thanks{faical.barzi@edu.uiz.ac.ma},  
 H.  El Moumni\orcidHasan\!\!$^1$\thanks{h.elmoumni@uiz.ac.ma (Corresponding author)}, K. Masmar\orcidKarima\!\!$^2$\thanks{karima.masmar@gmail.com}
\\
{\small $^{1}$ LPTHE, Physics Department, Faculty of Sciences, Ibnou Zohr University, Agadir, Morocco. }\\
{\small $^{2}$Laboratory of  High Energy Physics and Condensed Matter
HASSAN II University,}\\{\small Faculty of Sciences Ain Chock, Casablanca, Morocco.}\\
{\small $^{3}$CRMEF, Regional Center for Education and Training Professions Marrakesh, Morocco.
}
}

\date{\today}
\begin{document} 
\maketitle
\begin{abstract}
Motivated by the nonextensive nature of entropy in gravitational context and the Gauge/Gravity duality, the black hole  thermodynamics has been attracting intense emphasis in the literature. Along the present work, we investigate some features of the phase structure and critical phenomena of the 4-dimensional charged black holes in asymptotically flat spacetime within the formalism of Rényi statistics. First, we explore the extended phase space via the Rényi statistics approach. Concretely, based on the modified version of the Smarr formula, we recall the equal-area law to remove the oscillatory non-physical region in the $P_R-V$ and $T_R-S_R$ planes. Then, the coexistence curves are determined, as well as the latent heat of phase change. Moreover, we prove that the critical exponent describing the behavior of the order parameter near the critical point is $\frac{1}{2}$, which is consistent with Landau's theory of continuous phase transition. Lastly, we apply the Hamiltonian approach to Rényi thermodynamics which provides a new and solid mathematical basis for the extension of phase space and puts more insight into an expected and profound possible connection between the nonextensivity Rényi parameter $\lambda$ and the cosmological constant $ \Lambda$.
{\noindent}
\end{abstract}
\newpage
\tableofcontents



\section{Introduction}\label{sect1}
\paragraph{}Black holes are nowadays the leading astrophysical and theoretical laboratories for testing general relativity as well as theories of modified and possibly quantum gravity. Additionally, the most fascinating achievement in modern physics is the bridge between thermodynamics and gravity, which plays a key role to understand more deeply the nature of black holes\cite{Bardeen:1973gs,Hawking:1982dh}.  Especially, the thermodynamic aspects of Anti-de Sitter ($AdS$) spacetimes are largely  considered in the literature \cite{Banerjee:2011raa,Haldar:2018vqo,Lemos:2015zma,Myung:2013uka}, while considering cosmological constant
$\Lambda$ as a thermodynamical variable opens new windows in such area and new phenomena familiar from everyday thermodynamics emerge, such as enthalpy, reentrant phase transitions, triple points, and Carnot cycle have all now entered the language and structure of the subject, broadening it to what is called Black Hole Chemistry\cite{Kubiznak:2014zwa,Kubiznak:2016qmn,Karch:2015rpa,Gregory:2019dtq,Astefanesei:2019ehu,Dehghani:2019thq,Dehghani:2020blz,Dehghani:2020kbn,Kumar:2020cve}. Maybe the only intriguing exception to this similitude is the non-extensive nature of the Bekenstein-Hawking entropy of black holes which is proportional to the surface area of its event horizon rather than the volume. Furthermore, in the strong gravitational scheme and in the vicinity of black holes, the condition of negligible long-range type interactions in the standard statistical descriptions break down, and consequently, the usual definition of mass and other extensive quantities is not possible locally which pushes us to go beyond the standard Gibbs-Boltzmann ($GB$) statistical proposal.

In other words, the black hole entropy $S_{BH}$ should encode the black hole information with non-local and non-extensive nature. Hence, our requirement for a non-Boltzmannian statistics is crucial. Abe has introduced  in  \cite{Abe},  the most general non-additive entropy composition rule as
\begin{equation} 
H_{\lambda}\left(S_{12}\right)=H_{\lambda}\left(S_{1}\right)+H_{\lambda}\left(S_{2}\right)+\lambda H_{\lambda}\left(S_{1}\right) H_{\lambda}\left(S_{2}\right), \label{non additive}
\end{equation} 
in which $H_{\lambda}$ stands for a differentiable function of entropy $S$, $\lambda$ denotes a real constant parameter, and $S_1$, $S_2$ and $S_{12}$ are the entropies of the subsystems and the total system, respectively. Next, Bir\'o and V\'an  extend this formula by exploring non-homogenous systems as well \cite{BV} and elaborate a formalism to derive the most general functional form of those non-additive entropy composition rules that satisfy the familiar relations of standard thermodynamics, especially the zeroth law of thermodynamics
\begin{equation}\label{formlog}
L(S)=\frac{1}{\lambda}\ln[1+\lambda H_{\lambda}(S)],
\end{equation}
and the compatibility with the additive composition feature
\begin{equation}
L(S_{12})=L(S_{1})+L(S_{2}).
\end{equation}
Moreover, the  harmonious temperature function with related zeroth law reads as
\begin{equation}
\frac{1}{T}=\frac{\partial L(S(E))}{\partial E}, 
\end{equation}
under the assumption of energy composition additivity. Within this proposal,  Alfréd Rényi in 1959 gave a well-defined entropy function, which also obeys both the equilibrium and the zeroth law compatibility exigency of thermodynamics. 
According to \cite{Renyi:1959aa}, the R\'enyi entropy  is defined as 
\begin{equation}\label{SR}
S_R=\frac{1}{1-q}\ln\sum_ip^q_i. 
\end{equation}
It is obvious  that $H_{\lambda}(S)=S$ and $\lambda=1-q$ in Eq.\eqref{formlog}  if the original entropy functions  follow the non-additive composition rule 
\begin{equation}\label{tr}
S_{12}=S_1+S_2+\lambda S_1S_2. 
\end{equation}
This expression is known by the Tsallis composition rule, and $\lambda\in\mathbb{R}$ is the parameter of nonextensivity, and Tsallis entropy is defined as 
$S_T=\frac{1}{1-q}\sum_i(p^q_i-p_i)$ \cite{Tsallis1}. By recalling the formal logarithm, one can associate  the Tsallis formula to the R\'enyi entropy by
\begin{equation}\label{ll}
S_R \equiv L(S_T)=\frac{1}{1-q}\ln\left[1+(1-q)S_T\right].
\end{equation}
Recovering the standard Boltzmann-Gibbs entropy, $S_{BG}=-\sum p_i\ln p_i$ from Tsallis or R\'enyi formula is an easy task, which is ensured by taking the vanishing limit of $\lambda$,  ($\lambda \rightarrow 0\equiv q\rightarrow 1$).
 Recently, Rényi entropy evinces very interesting features in the black hole thermodynamics context and attracts a special emphasis in literature 
\cite{Mejrhit:2019oyi,Tannukij:2020njz,Promsiri:2020jga,Czinner:2015eyk,Czinner:2017tjq,dilaton,Promsiri:2021hhv,Hirunsirisawat:2022ovg,BARZI2022137378}. By recalling the Rényi statistics framework, both \cite{Promsiri:2020jga,dilaton} studies reveal that it is possible, with $0<\lambda<1$, to obtain the small and large black hole branches
in the grand canonical ensemble while this cannot occurs in
the $GB$ statistics approach. Moreover, they unveil that the Hawking–Page phase transition between thermal radiation and large black holes in asymptotically flat Reissner-Nordström (RN-flat) and in the presence of dilaton field can happen and crucially depend on the nonextensivity parameter $\lambda$. 
Nevertheless, in the canonical ensemble, the black hole exhibits a critical behavior such that a small/large black hole (SBH/LBH) first-order phase transition when $\lambda < \lambda_c$. Above the critical value of the Rényi parameter $\lambda_c$, this behavior disappears and the large black hole phase is the only remaining phase that is possible. Rigorously speaking, this thermal picture is, in the same way, as that of the charged and dilatonic black holes in $AdS$ space with the standard $GB$ statistics.
It is commonly known from black hole phase structure investigations that when the system reaches the requirement of thermodynamic equilibrium stability in an isothermal or isobaric process, the $P-V$ or $T-S$ curves describing the change of the system are discontinuous, and the system possesses a latent heat of phase transition when it crosses the two-phase coexistence line\cite{Guo:2019oad}. This discontinuity is associated with a thermodynamically unstable region when $\partial P/\partial V>0$ or $\partial T/\partial S<0$, this nonphysical situation is resolved by Maxwell's construction \cite{Spallucci:2013osa,Spallucci:2013jja,Belhaj:2014eha,Wei:2014qwa,Belhaj:2015hha}.
\paragraph{}Besides, Gauge/Gravity duality states a correspondence between gauge and gravity theories. Since the gauge theory side of the duality supports standard thermodynamic treatment, it is natural to demand that standard thermodynamics should be present on the gravity side of the duality. However, the minimal asymptotically Anti-de Sitter Reissner-Nordström (RN-AdS) has a one-dimensional thermodynamic description where the entropy is the only degree of freedom that doesn't match the multidimensional thermodynamical nature of most usual matter systems. The underlying reason is that in a formulation with one free variable it is impossible to draw a distinction between isentropic and isothermic thermodynamical transformations, and the fundamental notion of thermodynamic temperature is ill-defined. Therefore, by increasing the number of degrees of freedom of black hole systems one is led to thermodynamics closer to standard thermodynamics of matter systems. Then one may ask how to extend the black hole thermodynamics in a consistent way. One such prescription is the Hamiltonian approach \cite{Baldiotti:2016lcf,Baldiotti:2017ywq,Haldar:2019pwt} to thermodynamics. Indeed, investigating thermodynamics from the Hamiltonian point of view is the newest tool to probe the thermal black hole laws and to extend the phase space of black holes in a mathematically sound manner compatible with the laws of black hole thermodynamics. Henceforward promoting the Rényi formalism to the Hamiltonian approach is a decisive task.
\paragraph{}Motivated by those works, herein, we intend to study some aspects of phase equilibrium of charged asymptotically flat black holes within Rényi formalism. Specifically, we implement Maxwell's construction in the phase portrait of RN-flat black holes through the Rényi statistics to remove non-physical behaviors in complete analogy with the Van der Waals ($VdW$) fluid and the RN-AdS black holes in the Gibbs-Boltzmann formalism. Then, we unveil the thermodynamical and Hamiltonian consequences of this implementation.

\paragraph{}The outline of the paper is as follows: In Sec.\ref{sect2} we give a thermodynamical short review of the asymptotically flat-charged black hole in Rényi statistics. In Sec.\ref{sect3} we implement Maxwell's equal-area law for the RN-flat black hole within Rényi formalism, first in the $P_R-V$ diagram by replacing the oscillatory curves with isobaric plateaus, then we perform the construction in the $T_R-S_R$ diagram where this time isothermic plateaus supplant the nonphysical trait of the phase profile. Based on the results of this section, in Sec.\ref{sect4} we derive the coexistence curves $P_0-T_0$ and the associated latent heat of first-order phase transition. Then, we analyze the influence of each parameter on the latent heat and provide a microscopic interpretation of the phase transition through the application of Landau's theory of continuous phase transition and its symmetry change argument. In Sec.\ref{sect5}, we apply the Hamiltonian approach to the RN-flat black hole in Rényi thermodynamics to consistently extend its phase space. Finally, Sec.\ref{sect6} is devoted to the conclusion.

\section{Charged black hole thermodynamics in Rényi statistics}\label{sect2}

The line element, describing the Reissner-Nordstrom black hole solution of mass $M$ and charge $Q$ can be written as
\begin{eqnarray}
ds^2 = -f(r)dt^2 + \frac{dr^2}{f(r)} + r^2(d\theta^2 + \sin^2\theta d\phi^2),
\end{eqnarray}
the  involved blackening function $f(r)$ reads as
\begin{eqnarray}
f(r)=1-\frac{2M}{r} + \frac{Q^2}{r^2}.
\end{eqnarray}
 The horizon radius $r_h$ is determined as
the largest real root of $f(r)\big|_{r=r_h} = 0$ and leads to the following mass formula:
\begin{eqnarray}
M = \frac{r_h}{2}\left( 1 + \frac{Q^2}{r_h^2} \right), \label{mass}
\end{eqnarray}
and the electric potential expression is obtained to be
\begin{equation}
\Phi = \frac{Q}{r_{h}} = \frac{Q}{M + \sqrt{M^2 - Q^2}}. \label {bh6}
\end{equation}

 According to \cite{Promsiri:2020jga,dilaton}, the  R\'enyi entropy $S_R$ is the formal logarithm of Bekenstein-Hawking one $S_{BH}$ taken as the Tsallis entropy $S_{T}$ in Eq.\eqref{ll} such as, 
\begin{equation}
S_R=\frac{1}{\lambda}\ln(1+\lambda S_{BH}). \label{bh17}
\end{equation}
 It's worth noting that in the limit of vanishing nonextensivity parameter $\lambda$, we recover the usual Gibbs-Boltzmann statistics, $S_R\overset{\lambda\rightarrow0}{\longrightarrow}S_{BH}$. In the present paper we assume that $\lambda$ is small and positive, $0<\lambda<<1$, guided by the assumption that non-extensive effects are first order corrections to the classical extensive statistical mechanics. 
\paragraph{} The R\'enyi entropy generalizes the Gibbs-Boltzmann entropy by taking into account non-extensive effects, this implies a correction to first order in $\lambda$ to the RN-flat black hole temperature. The Rényi temperature $T_R$ can be expressed as \cite{Promsiri:2020jga},
\begin{eqnarray}
T_R = \frac{1}{\partial{S_R/\partial{M}}} &=& T_H(1+\lambda S_{BH})\label{Tr}\\ 
&=& \frac{(r_h^{2}-Q^2)(1+ \lambda \pi r_h^{2})}{4\pi r_h^{3}}. \label{bh25}
\end{eqnarray}
 Where $T_H= \frac{r_h^{2}-Q^2}{4\pi r_h^{3}}$ and $S_{BH}=\pi r_h^{2}$ are the Hawking temperature and the Bekenstein-Hawking entropy of RN-flat black hole, respectively.
Moreover, in the Rényi extended phase space\cite{Promsiri:2020jga}, the nonextensivity parameter $\lambda$ is associated with the Rényi thermodynamic pressure and the electric potential $\Phi$ by
\begin{eqnarray}
P_R = \frac{3\lambda (1-\Phi^2)}{32}.
\label{p-lam}
\end{eqnarray}
In this framework, the conjugate quantity associated with the pressure is the thermodynamical volume $V=\frac{4}{3}\pi r_h^3$, this assertion will be made mathematically more precise in section \ref{sect5}. Summarizing all these quantities, one can write the first law of Rényi thermodynamics and its associated generalized Smarr formula as,
\begin{eqnarray}
dM &=& T_RdS_R + VdP_R + \Phi dQ. 
\\
M &=& 2T_RS_R - 2P_RV + \Phi Q. \label{Smarr_rényi_mod}
\end{eqnarray}
Otherwise, one may assume that the black holes could have their own micro-states carrying the degrees of freedom as well and the specific volume $v$, which is proportional to the event horizon radius \cite{Promsiri:2020jga}
\begin{equation}\label{v_rh}
v=\frac{4}{3}r_h.
\end{equation}

 Having presented some thermodynamical features of the RN-flat black hole including the thermodynamical quantities associated with Rényi statistics, we pave the way for the study of the corresponding phase structure and the Maxwell equal areas law in such black hole background via this non-Boltzmannian formalism.
 

\section{The equal-area law of the asymptotically flat charged black hole in Rényi formalism}\label{sect3}
  In this section, we examine Maxwell's construction of the charged flat black hole. To do this, we must first recall that Maxwell developed the equal area law to describe the experimental behaviors of the state equations associated with real fluids.
The (Pressure, Volume)-plane is often where this design is elaborated while maintaining a constant temperature.
Such a construction can, however, also be made in the (Temperature, Entropy) plane by fixing the pressure this time. \\

To derive the equation of state in the $P_R-V$ diagram, first we invert the Rényi temperature expression given by the
 Eq.\eqref{bh25}  and in where we substitute $\lambda$ by Eq.\eqref{p-lam} and the electric potential $\Phi$ by its form Eq.\eqref{bh6}, we get  
\begin{equation}\label{bhr1}
P_R=\displaystyle    \frac{3 T_{R}}{8 r_{h}}- \frac{3}{32 \pi r_{h}^{2}}+\frac{3 Q^{2}}{32 \pi r_{h}^{4}}.
\end{equation}
Now through the expressions of the specific volume, $v=\frac{8}{3}r_h$ and the thermodynamical volume, $V=\frac{4}{3}r_h^3$ in terms of $r_h$ we arrive at the desired equations of state
\begin{equation}\label{bhr2}
P_R=  \frac{T_R}{v}  - \frac{2}{3 \pi v^{2}} + \frac{128 Q^{2}}{27 \pi v^{4}} \quad \Leftrightarrow\quad P_R=\displaystyle  \frac{6^{\frac{2}{3}} \sqrt[3]{\pi} T_{R}}{8 {V}^{\frac{1}{3}}}- \frac{\sqrt[3]{6}}{16 \sqrt[3]{\pi} V^{\frac{2}{3}}}+\frac{6^{\frac{2}{3}} \sqrt[3]{\pi} Q^{2}}{24 V^{\frac{4}{3}}}.
\end{equation}

\begin{figure}[!ht]
\centering
	\includegraphics[scale=.55]{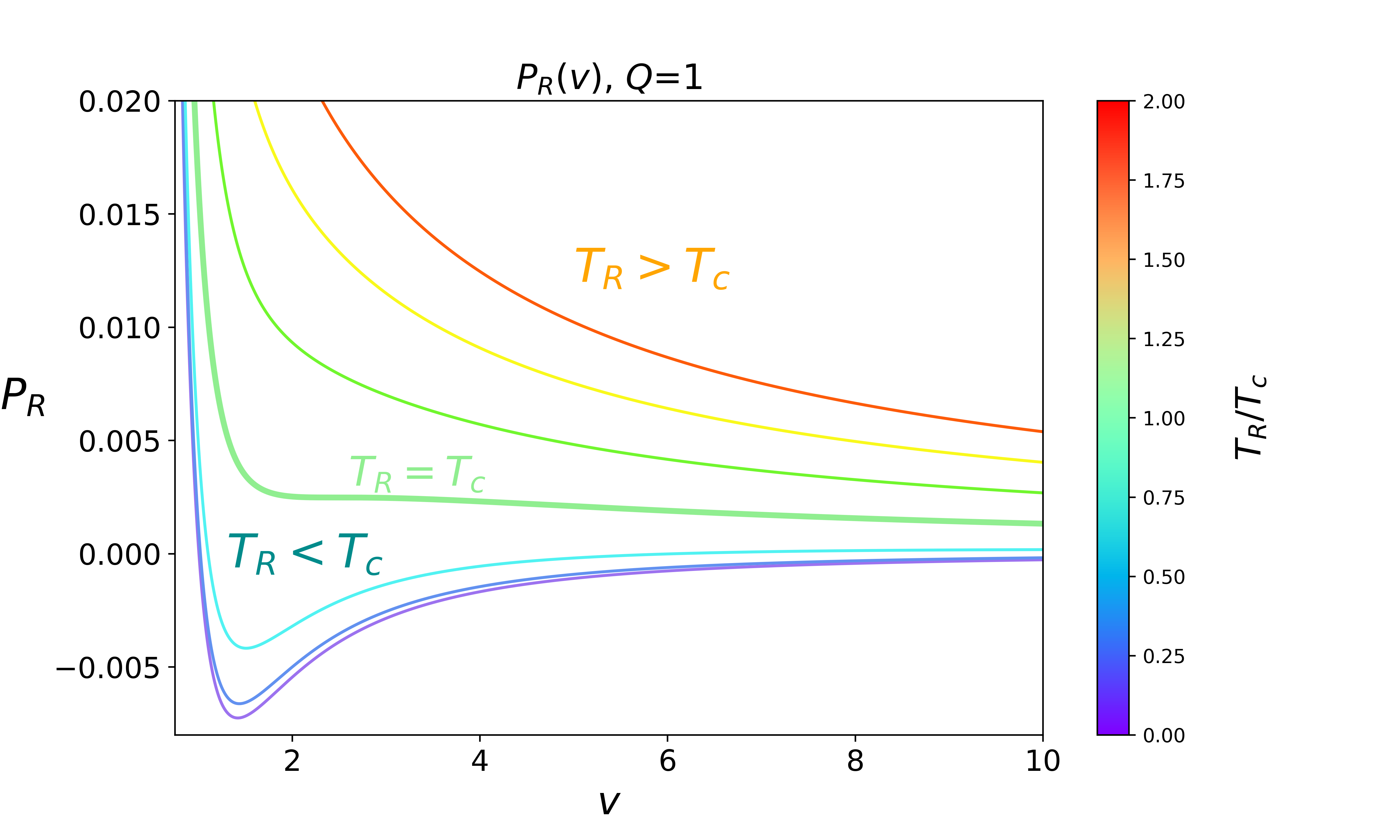}
	\vspace{-.5cm}
	\caption{\footnotesize {\it Isotherms of the charged asymptotically flat black hole in $P_R-v$ diagram with an electric charge $Q=1$ in Rényi thermodynamics. The green thick curve is the critical isotherm at $T_R=T_c$,  It is observed that below the critical isotherm the unphysical oscillating behavior   appears. }}\label{fig:p-v}
\end{figure}

 The first form of the equation of state associated with the charged asymptotically flat black hole in the $P-v$ diagram is exhibited by the isotherms in Fig.\ref{fig:p-v}, 
in which the thermodynamically unstable state associated with $\partial P_R/\partial v>0$ will lead to the system automatically expand or contract unrestrictedly. Such a situation occurs also in the Van der Waals ($VdW$) equation and in a variety of black holes configurations in the extended phase space \cite{Spallucci:2013osa,Spallucci:2013jja,Zhang:2014fsa,Belhaj:2014eha,Miao:2016ieh,Nguyen:2015wfa}, but it has been fixed by recalling Maxwell's equal-area law.  Moreover, Fig.\ref{fig:p-v} exhibits a critical behavior where the black hole undergoes a second phase transition at a critical temperature $T_c=\frac{\sqrt{6}}{18 \pi Q}$ whose analytical expression can be derived by solving the following system
  \begin{equation}
 \left(\frac{\partial P_R}{\partial v}\right)_{T_R,Q}=
 \left(\frac{\partial^2 P_R}{\partial v^2}\right)_{T_R,Q}=0.
  \end{equation}


In what follows, we extend Maxwell's construction formalism to incorporate Rényi statistics by studying charged asymptotically flat back hole phase structure.

\subsection{The construction of equal-area law in $P_R-V$ diagram}
\begin{figure}[!ht]
		 \centering		
			\begin{tabbing}
			\centering
			\hspace{-1.3cm}
			\includegraphics[scale=0.38]{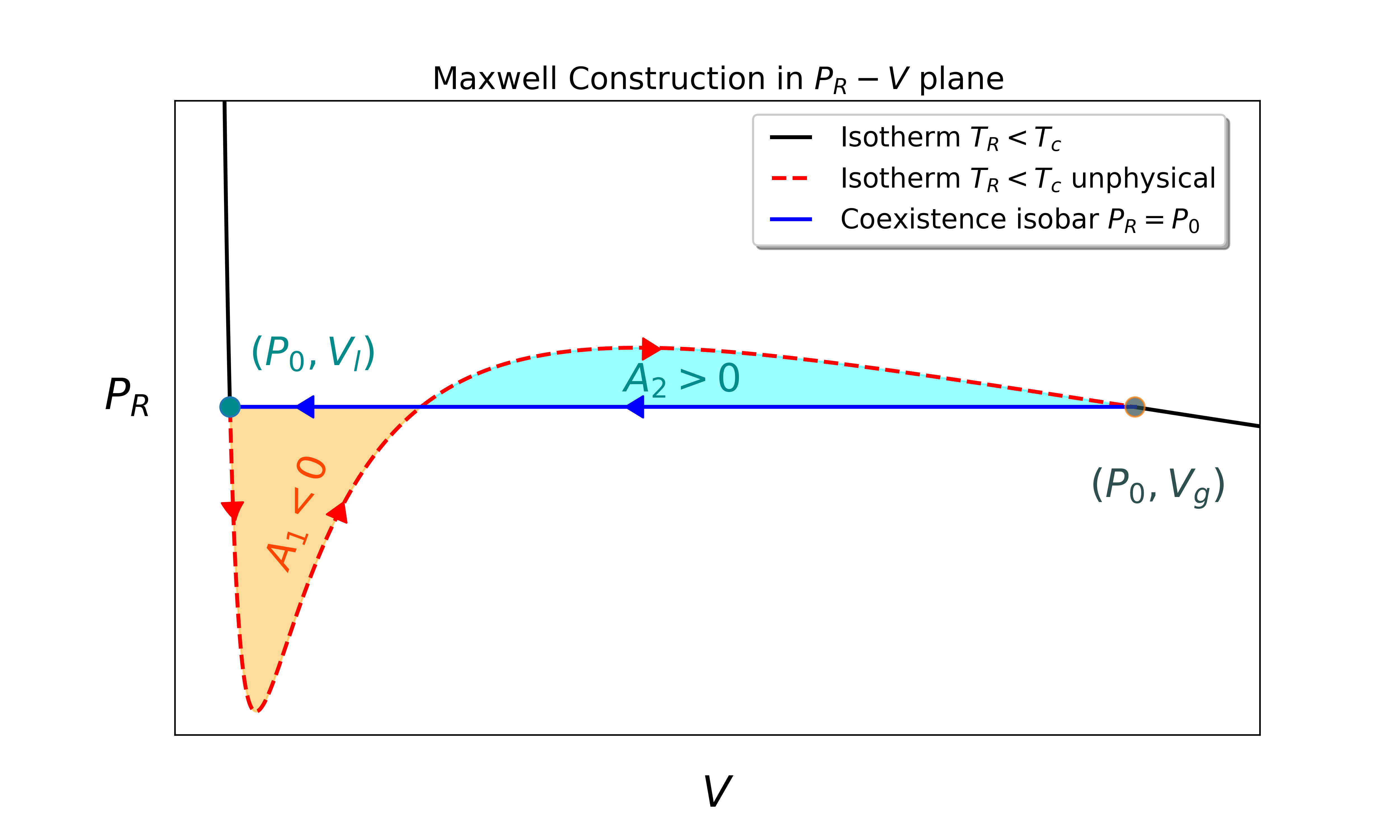}
            \hspace{-1.cm}
			\includegraphics[scale=.38]{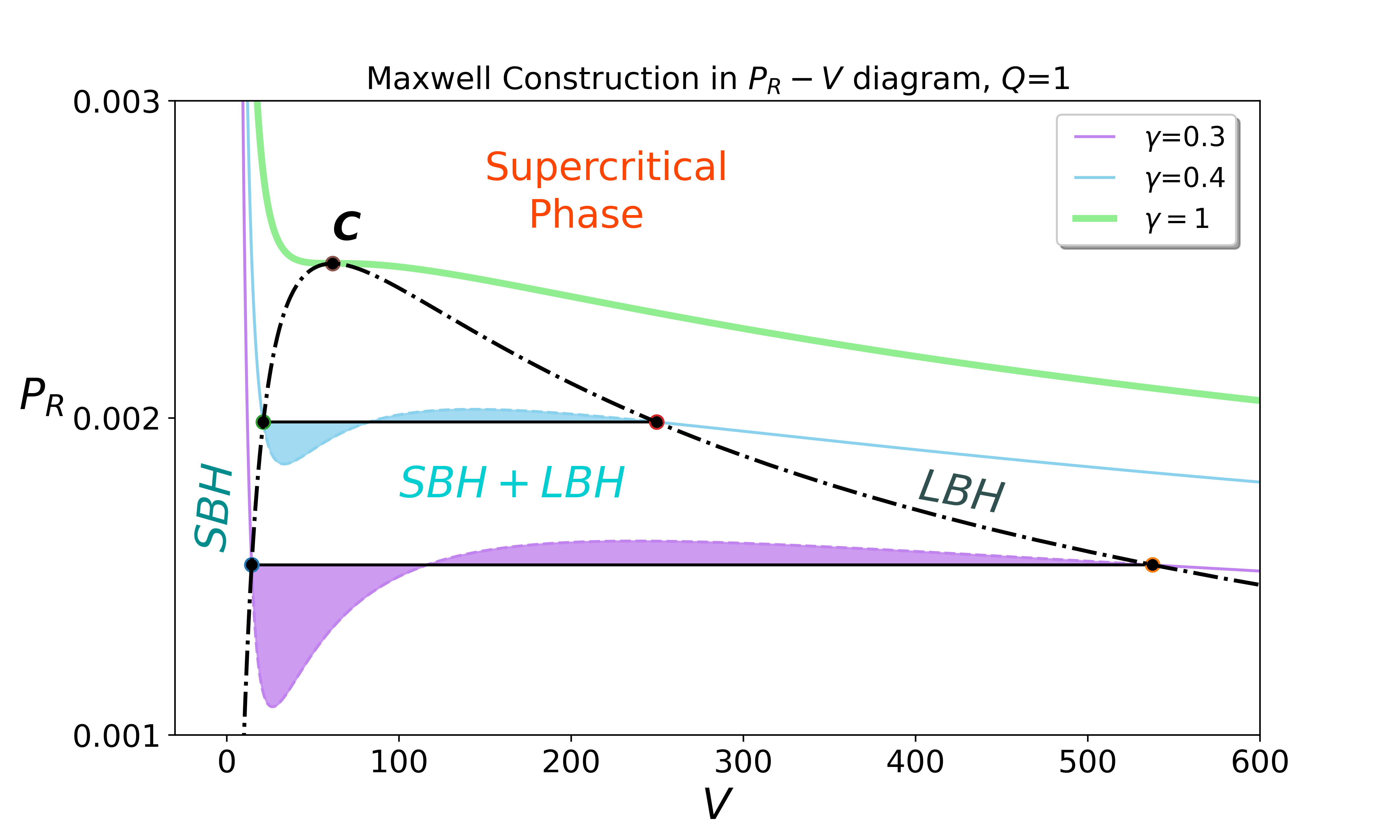} 
			\end{tabbing}
			\vspace{-1.cm}
 	      \caption{\footnotesize{\it The simulated phase transition and the boundary of the two phase coexistence on the base of the isotherm in the  $P_R-V$ diagram for charged black hole in flat spacetime within Rényi statistics approach. \textbf{Left panel:} demonstration of the Maxwell construction in the $P_R-V$ plan, the blue thick line is calculated such as the two shaded areas are equal and to eliminate the unphysical behavior represented by the red dotted line. \textbf{Right panel:} the black horizontal lines  are isotherms replacing the unphysical oscillations, the bell shaped black dashed line delimits the coexistence region. The critical isotherm is shown as the thick green line above which the supercritical phase dominates. 
	    }}
	\label{fig:mc_p_V}
\end{figure}

\paragraph{}Maxwell's construction in the $P_R-V$ plane is based on the property of Helmholtz free energy $F_R$ being a state function of the black hole. In the left panel of Fig.\ref{fig:mc_p_V}, the black hole undergoes a reversible cyclic transformation along an isotherm curve $T_R<T_c$, 
starting from the liquid state point $(P_0,V_l)$ to the gaz state point $(P_0,V_g)$ along the red dashed curve and coming back along the blue line, we have
\begin{equation}\label{bhr_int_F}
\oint dF_R=0.
\end{equation}
Where the differential of $F_R$ at constant charge $Q$ is given by
\begin{equation}\label{bhr_dF}
dF_R=-S_RdT_R-P_RdV.
\end{equation}
Along the red dashed curve, the differential of $F_R$ takes the form $dF_R=-P_RdV$, while on the blue line $P_R=P_0$, it reduces to $dF_R=-P_0dV$. Thus, from Eq.\eqref{bhr_int_F} we write,
\begin{equation}\label{bhr_F}
\oint dF_R = -\int_{V_l}^{V_g}P_R\:dV -\int_{V_g}^{V_l} P_0\:dV=0.
\end{equation}
Which gives the form of  the Maxwell's equal-area law in the $P_R-V$ diagram as 
\begin{equation}\label{bhr_maxwell_PV}
P_0(V_g-V_l)= \int_{V_l}^{V_g} P_R\:dV,
\end{equation}
and traduces the equality of the areas delimited by the blue line and the dashed red curve shown in the left panel of Fig.\ref{fig:mc_p_V}.

 \paragraph{}In the right panel of  Fig.\ref{fig:mc_p_V} we have plotted the isotherms in the $P_R-v$ plane according to the equation of state Eq.\eqref{bhr2} for a charge $Q=1$. We can see clearly from this plot the thermodynamically unstable regions below the critical isotherm $T=T_c$ where the compressibility of the black hole phase is negative $(\partial P_R/\partial V>0)$, this behavior being unphysical,  the construction of Maxwell aims for replacing these regions by horizontal lines where temperature and pressure are constants. Moreover, below a given temperature, negative pressure appears, although Maxwell's construction can remove some negative pressure zones, others will persist as seen from Fig.\ref{fig:p-v} for very low temperatures compared to the critical temperature $T_c$. For a $VdW$ fluid, negative pressure is associated with metastable states of the liquid phase, while in our case, these regions of instability can be viewed as metastable states where the black hole is stretched under a background tension. 
 
  \paragraph{}Keeping the charge $Q$ constant and  on an arbitrary isotherm associated with $T_R=T_0\leq T_c$, the two points $(P_0,V_l)$ and  $(P_0,V_g)$ meet the  Maxwell's equal-area law Eq.\eqref{bhr_maxwell_PV} :
\begin{equation}\label{bhr4}
P_0(V_g-V_l)= \int_{V_l}^{V_g} P_R\:dV=\int_{r_l}^{r_g} 4\pi r_h^2 P_R(r_h)\:dr_h.
\end{equation}
Where $V_l=\frac{4\pi}{3}r_l^3$ associated with the small black hole phase and $V_g=\frac{4\pi}{3}r_g^3$ labeling the large black hole one.  Integrating with respect to $r_h$ for convenience, we obtain the coexistence pressure $P_0$ as,
\begin{equation}\label{bhr5}
P_0= \frac{9 \left(Q^{2} + 2 \pi T_{0} r_{g}^{2} r_{l} + 2 \pi T_{0} r_{g} r_{l}^{2} - r_{g} r_{l}\right)}{32 \pi r_{g} r_{l} \left(r_{g}^{2} + r_{g} r_{l} + r_{l}^{2}\right)}.
\end{equation}
From Eq.\eqref{bhr1}, we obtain for each mentioned state
\begin{equation}\label{bhr30}
P_0=\displaystyle \frac{3 Q^{2}}{32 \pi r_{l}^{4}} - \frac{3}{32 \pi r_{l}^{2}} + \frac{3 T_{0}}{8 r_{l}},
\end{equation}
\begin{equation}\label{bhr31}
P_0=\displaystyle \frac{3 Q^{2}}{32 \pi r_{g}^{4}} - \frac{3}{32 \pi r_{g}^{2}} + \frac{3 T_{0}}{8 r_{g}}.
\end{equation}
Summing Eqs.\eqref{bhr30} and \eqref{bhr31} we get
\begin{equation}\label{bhr32}
 2 P_{0} = \frac{3 Q^{2}}{32 \pi r_{l}^{4}} + \frac{3 Q^{2}}{32 \pi r_{g}^{4}} + \frac{3 T_{0}}{8 r_{l}}-\frac{3}{32 \pi r_{l}^{2}} - \frac{3}{32 \pi r_{g}^{2}} + \frac{3 T_{0}}{8 r_{g}} 
\end{equation}
\begin{equation}\label{bhr33}
   \implies \displaystyle 2 P_{0} = \frac{3 Q^{2}}{32 \pi r_{g}^{4}(1+\gamma^{4})} - \frac{3}{32 \pi r_{g}^{2}(1+\gamma^{2})} + \frac{3 T_{0}}{8 r_{g}(1+\gamma)}.   
\end{equation}

Where we have introduced the ratio $\gamma=\frac{r_l}{r_g}$. Furthermore, the subtraction of Eqs.\eqref{bhr30} and \eqref{bhr31} gives
\begin{equation}\label{bhr6}
\displaystyle 0=T_0- \frac{ Q^{2}( 1-\gamma^{4})  + \gamma^{2} r_{g}^{2}(\gamma^{2}-1)}{4 \pi \gamma^{3} r_{g}^{3} \left(\gamma - 1\right)}.
\end{equation}\\
While, from Eq.\eqref{bhr5} and by substituting $r_l$ by $\gamma r_g$ one can obtain
\begin{equation}\label{bhr7}
0=\displaystyle 18 \pi T_{0} \gamma r_{g}^{3} \left( \gamma + 1\right)- 32 \pi P_{0} \gamma r_{g}^{4} \left(\gamma^{2} + \gamma + 1\right)  - 9 \gamma r_{g}^{2}+ 9 Q^{2}. 
\end{equation}
Next, injecting Eqs.\eqref{bhr33} and \eqref{bhr6} into Eq.\eqref{bhr7} we get,
\begin{equation}\label{bhr34}
0=\displaystyle r_{g}^{2} \left(\gamma^{6} - 4 \gamma^{5} + 6 \gamma^{4} - 4 \gamma^{3} + \gamma^{2}\right)+Q^{2} \left(- \gamma^{6} + 9 \gamma^{4} - 16 \gamma^{3} + 9 \gamma^{2} - 1\right). 
\end{equation}
Solving Eq.\eqref{bhr34} for $r_g$ we obtain
\begin{equation}\label{bhr9}
r_g(\gamma)= \frac{Q \sqrt{\gamma^{2} + 4 \gamma + 1}}{\gamma},
\end{equation}
and thus leading also to $r_l$,
\begin{equation}\label{bhr10}
r_l(\gamma)= Q \sqrt{\gamma^{2} + 4 \gamma + 1}.
\end{equation}  Consequently, we get the volume of each state as
\begin{equation}
V_g(\gamma)=\displaystyle \frac{4 \pi Q^{3} \left(\gamma^{2} + 4 \gamma + 1\right)^{\frac{3}{2}}}{3 \gamma^{3}} \quad\text{ and }\quad V_l(\gamma)=\displaystyle \frac{4 \pi Q^{3} \left(\gamma^{2} + 4 \gamma + 1\right)^{\frac{3}{2}}}{3}.
\end{equation}
Advancing, by recalling Eq.\eqref{bhr31}, we derive the expression of $P_0$ in terms of the ratio $\gamma$ as
\begin{equation}\label{bhr12}
 P_0(\gamma)= \frac{9 \gamma^{2}}{32 \pi Q^{2} \left(\gamma^{2} + 4 \gamma +  1\right)^2} .
\end{equation}
At this level, one can notice that the critical point is attained in the limit $\gamma\longrightarrow 1$, thus  the critical radius $r_c$,  pressure $P_c$, and the volume $V_c$ are found to be :\begin{equation}\label{bhr11}
r_c=\sqrt{6}Q, \quad P_c= \frac{1}{128 \pi Q^{2}},\text{ and }\quad V_c=8\sqrt{6}\pi Q^3.
\end{equation}
These critical quantities are consistent with those in \cite{Promsiri:2020jga}.
Maxwell construction in the $P_R-V$ diagram is depicted in right panel of Fig.\ref{fig:mc_p_V} for different values of ratio $\gamma$.

%
%

In such panel, the bell shaped dashed black line is nothing but the so-called saturation line, and  $V_{l,g}$ present the intersection of this curve with an isotherm curve determined for each value of the parameter $\gamma$. Between these two values $V_l < V < V_g$, the black hole system is unstable; there is a phase transition between a small and a large black hole.  Thus, the oscillating part, below the critical pressure $P_c$, should be replaced with the isobar $P_R=P_0$ in the same way as it is done in the $VdW$ and RN-AdS cases and which reflects that $A_1$(orange), the colored area under the black solid line equals in an algebraic manner $A_2$(cyan), the colored area above the black solid line. This isobar $(P_R=P_0)$ is the locus of the coexistence of the two black hole phases, the size of this coexistence region decreases as the ratio $\gamma$ increases towards the critical value $\gamma_c=1$ where it vanishes. It is worth noting that the intersection points meet and correspond to the critical horizon volume/radius. This line represents all the possible critical points; therefore all possible phase transitions. While on the critical point $({\bm C})$, the two-phase system undergoes a second-order phase transition, a first-order phase transition occurs in all intersection points below $({\bm C})$. At the critical pressure $P_c$ and above the small and large black hole phases converge to a single supercritical phase, in this phase the two black holes (SBH and LBH) are indistinguishable.

\paragraph{}Inverting Eq.\eqref{bhr12} we can obtain $\gamma$, $r_l/V_l$ and $r_g/V_g$ directly in terms of $Q$ and the dimensionless quantity $\eta=Q\sqrt{P_0}$,

\begin{equation}\label{subs1}
\gamma=\frac{3-\sqrt{96 \eta  \left(4 \pi  \eta -\sqrt{2 \pi }\right)+9}}{8 \sqrt{2 \pi } \eta }-2
\end{equation}
\begin{equation}\label{bhr16}
r_l=\frac{Q\sqrt{9-48 \sqrt{2 \pi } \eta -3 \sqrt{96 \eta  \left(4 \pi  \eta -\sqrt{2 \pi }\right)+9}}}{8 \sqrt{\pi } \eta },
\end{equation}

\begin{equation}\label{bhr15}
r_g=\frac{\sqrt{6} Q}{\sqrt{3-16 \sqrt{2 \pi } \eta -\sqrt{96 \eta  \left(4 \pi  \eta -\sqrt{2 \pi }\right)+9}}}.
\end{equation}
And finaly,
\begin{equation}\label{bhr_Vl}
V_l=\frac{Q^3\left(9-48 \sqrt{2 \pi } \eta -3 \sqrt{96 \eta  \left(4 \pi  \eta -\sqrt{2 \pi }\right)+9}\right)^{3/2} }{384 \sqrt{\pi } \eta ^3},
\end{equation}

\begin{equation}\label{bhr_Vg}
\hspace*{-1.cm}
V_g=\frac{8 \sqrt{6} \pi  Q^3}{\left(3-16 \sqrt{2 \pi } \eta -\sqrt{96 \eta  \left(4 \pi  \eta -\sqrt{2 \pi }\right)+9}\right)^{3/2}}.
\end{equation}
Having established Maxwell's equal-area law construction associated which a charged-flat black hole in the (Pressure, Volume)-diagram within the Rényi statics framework, we will turn our attention in the next section to the (Temperature, Entropy)-plane.

\subsection{The construction of equal-area law in $T_R-S_R$ diagram}

\begin{figure}[!ht]
     \centering
	\includegraphics[scale=.55]{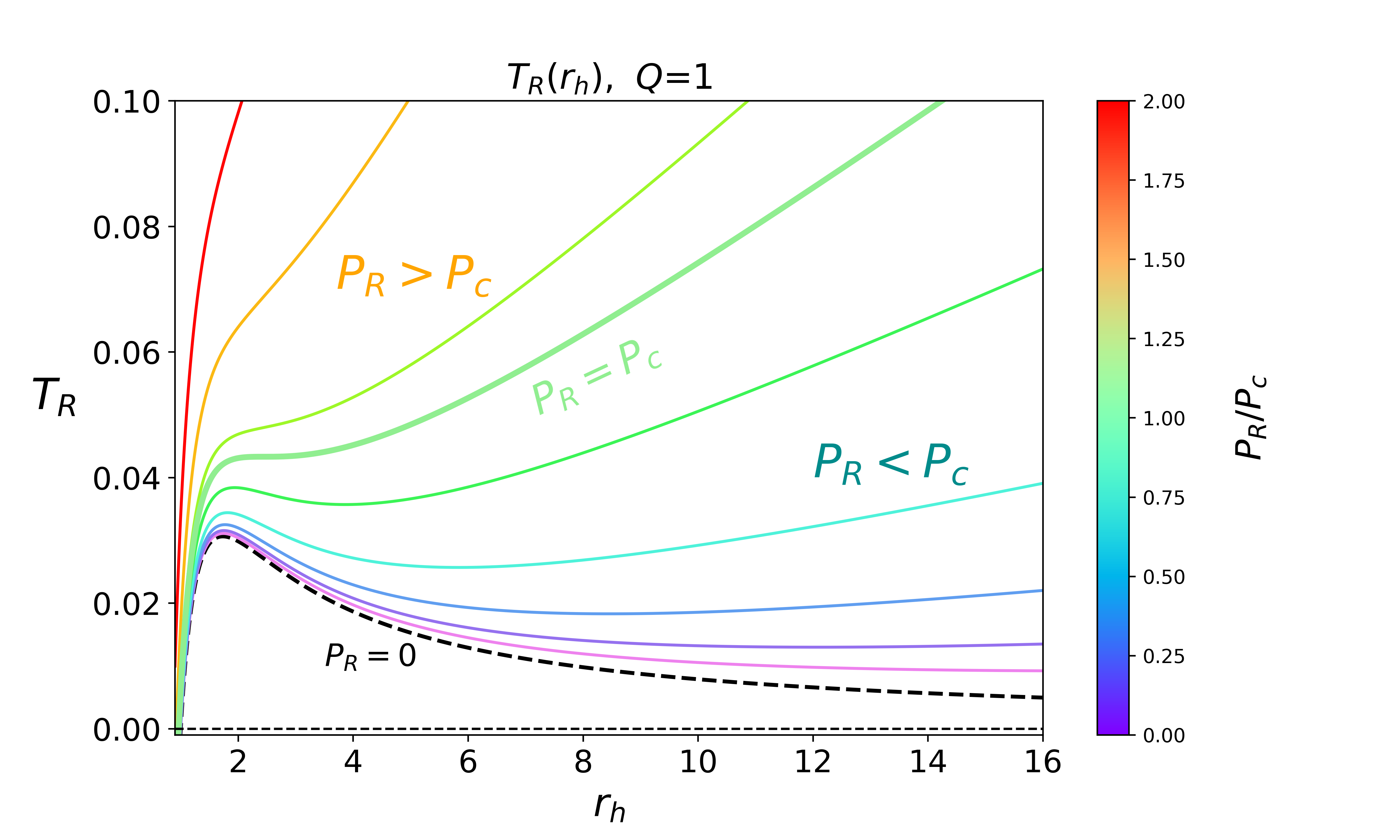}
	\vspace{-0.5cm}
	\caption{\footnotesize {\it Isobars  of the charged asymptotically flat black hole in $T_R-r_h$ plane with electric charge $Q=1$ in Rényi thermodynamics. The green thick curve is the critical isobar at $P_R=P_c$, It is observed that below the critical isobar the unphysical behavior appears and is represented by the negative slops of the isobars.}}
	\label{fig:cm_isobar_T_r}
\end{figure}

Herein, the equation of state Eq.\eqref{bhr1} solved for $T_R$  traduces the Rényi temperature variation in term of the event horizon radius and it reads as
\begin{equation}\label{bhr17}
T_R=\displaystyle   \frac{1}{4 \pi r_{h}}- \frac{Q^{2}}{4 \pi r_{h}^{3}} + \frac{8 P_R r_{h}}{3},
\end{equation}
and their associated  thermal profile $T_R-r_h$ curves are shown in Fig.\ref{fig:cm_isobar_T_r}.

 It's obvious that the Van der Waals-like phase transition persists across the oscillating region where the instability is represented by the negative slop region $(\partial T_R/\partial r_h<0)$ which corresponds to a negative heat capacity $(\partial S_R/\partial T_R<0)$, since $(\partial S_R/\partial r_h)$ is always positive. Also the critical behavior at the pressure $P_c$ which is given by Eq.\eqref{bhr11} implies a second order phase transition.

\begin{figure}[!ht]
	\centering
		\begin{tabbing}
			\centering
			\hspace{-1.3cm}
			\includegraphics[scale=.38]{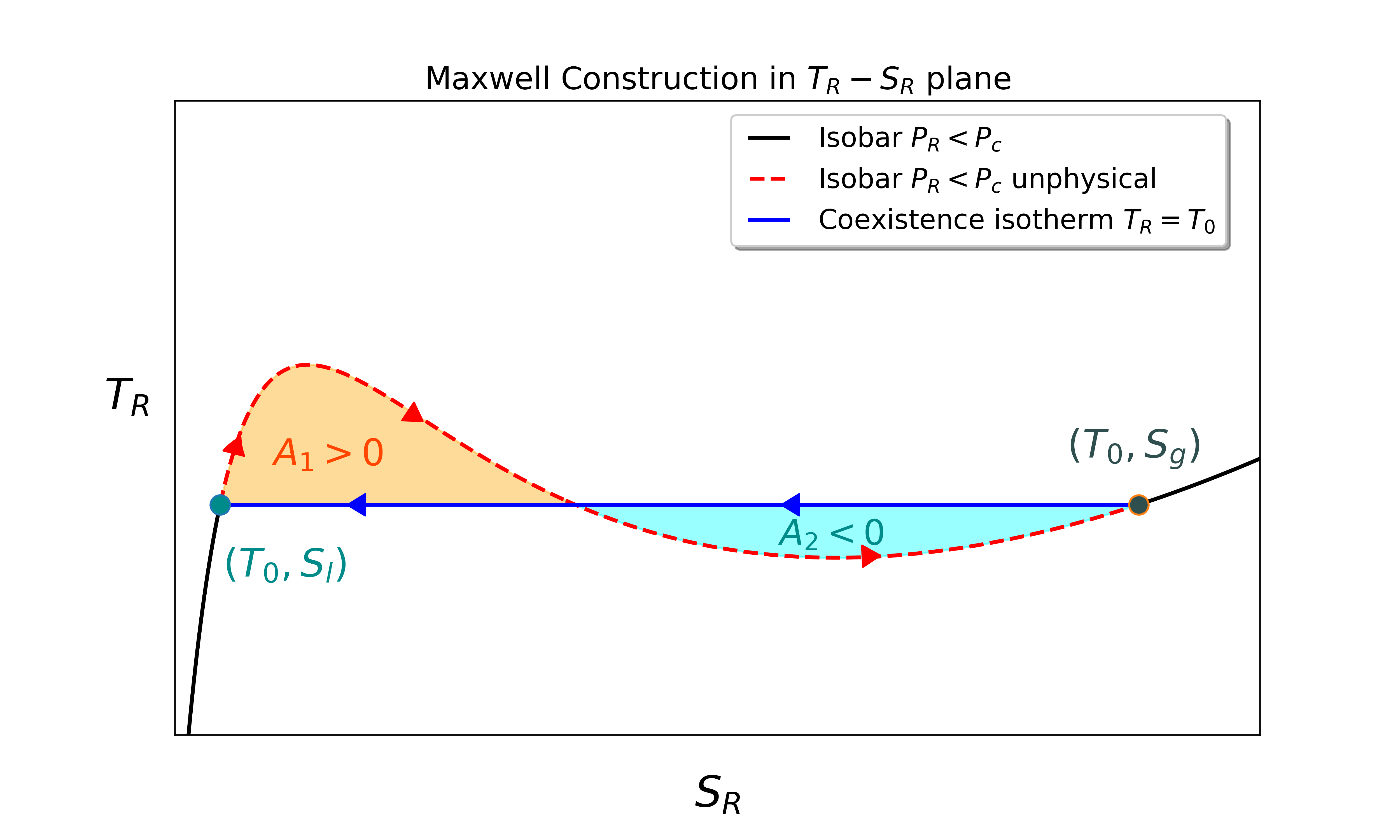}\label{}
            \hspace{-0.9cm}
	\includegraphics[scale=.38]{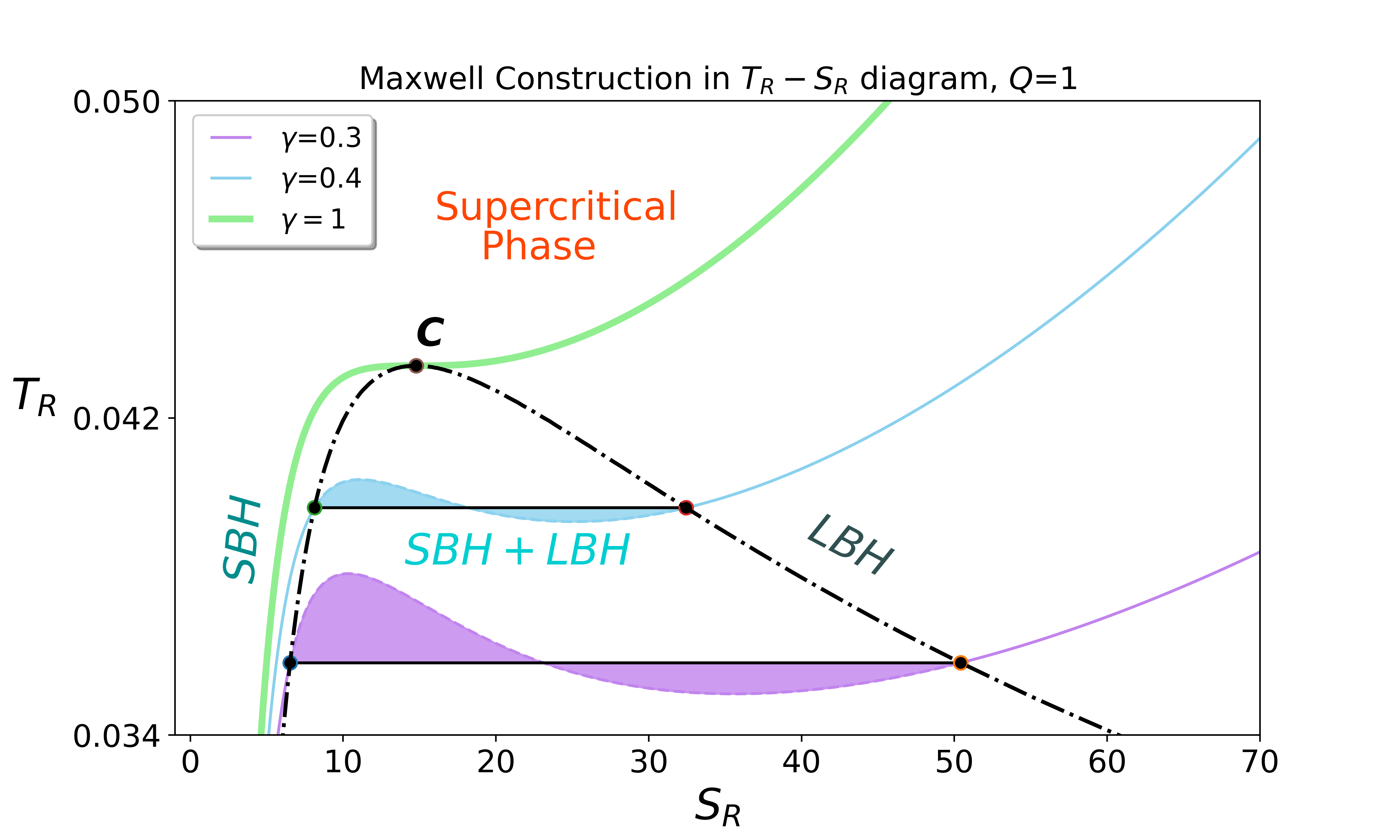}
	\end{tabbing}
	\vspace{-1.cm}
    \caption{\footnotesize {\it The simulated phase transition and the boundary of the two phase coexistence on the base of the isobar in the  $T_R-S_R$ diagram for charged black hole in flat spacetime within Rényi statistics approach. \textbf{Left panel:} demonstration of the Maxwell construction in the $T_R-S_R$ plan, the blue thick line is calculated such as the two shaded areas are equal and to eliminate the unphysical oscillatory behavior represented by the red dotted line. \textbf{Right panel:} black horizontal lines are isobars replacing the unphysical oscillations, the bell shaped black dashed line delimits the coexistence region. The critical isobar is shown as the thick green line above which the supercritical phase dominates.}}
    \label{fig:mc_T_S}
\end{figure}

\paragraph{}In the $T_R-S_R$ plane, a similar Maxwell's construction is done by replacing the Helmholtz energy by the Gibbs free energy $G_R$ as the function of state of the black hole.
As illustrated in  left panel of Fig.\ref{fig:mc_T_S}, the black hole undergoes a reversible cyclic transformation, we write
\begin{equation}\label{bhr_int_G}
\oint dG_R=0.
\end{equation}
Where the differential of $G_R$ at constant charge $Q$ is found to be
\begin{equation}\label{bhr_dG}
dG_R=VdP_R-S_RdT_R.
\end{equation}\\
On the red dashed isobar curve, the differential of $G_R$ reduces to $dG_R=-S_RdT_R$, while at the blue line $(T_R=T_0)$, it vanishes, $dG_R=0$. Thus, from Eq.\eqref{bhr_int_G} one can write
\begin{equation}\label{bhr_G}
\oint dG_R = -\int_{S_l}^{S_g}S_R\:dT_R =0.
\end{equation}
Integrating by part gives
\begin{equation}\label{bhr_G2}
\Big[ T_RS_R\Big]_{S_l}^{S_g}- \int_{S_l}^{S_g}T_R\:dS_R =0,
\end{equation}
which leads to the Maxwell's equal-area law form in the $T_R-S_R$ diagram and reflects the equality of the areas under the blue line and the dashed red curve of Fig.\ref{fig:mc_T_S},

\begin{eqnarray}\label{bhr18}\nonumber
T_0(S_g-S_l)&=& \int_{S_l}^{S_g} T_R\:dS_R\\&=&\int_{r_l}^{r_g} T_R\:\frac{dS_R}{dr_h}\:dr_h,
\end{eqnarray}
with $S_l=S_R(r_l)$ and $S_g=S_R(r_g)$. Recalling that $S_R=\displaystyle \pi r_{h}^{2} - \frac{\lambda \pi^{2}  r_{h}^{4}}{2} + \mathcal{O}\left(\lambda^{2}\right)$ and replacing $r_l$ and $r_g$ by Eq.\eqref{bhr16} and \eqref{bhr15} respectively,  we obtain for the coexistence temperature $T_0$ in terms of $\eta=Q\sqrt{P_0}$,
\begin{equation}
T_0=\displaystyle \frac{\sqrt{64 \pi  \eta ^2-16 \sqrt{2 \pi } \eta +\frac{3}{2}} \sqrt{-16 \sqrt{2 \pi } \eta -\sqrt{96 \eta  \left(4 \pi  \eta -\sqrt{2 \pi }\right)+9}+3}}{3 \pi  Q\left(\displaystyle\frac{\sqrt{96 \eta  \left(4 \pi  \eta -\sqrt{2 \pi }\right)+9}-3}{8 \sqrt{2 \pi } \eta }+3\right)}.
\end{equation}
In terms of the parameter $\gamma$, the coexistence temperature $T_0$ possesses a minimal expression  and becomes
\begin{equation}\label{bhr20}
T_0(\gamma)= \frac{\gamma \left(\gamma + 1\right)}{\pi Q \left(\gamma^{2} + 4 \gamma + 1\right)^{\frac{3}{2}}}.
\end{equation}
We give also the entropies of the small black hole phase $S_l$ and the large black hole one $S_g$ as
\begin{equation}
  S_l=\displaystyle \frac{\pi Q^{2} \left(\gamma + 4\right) \left(\gamma^{2} +  4\gamma + 1\right)}{3 \gamma } \ln{\left(\frac{4 \left(\gamma + 1\right)}{\gamma + 4}\right)},
\end{equation}
\begin{equation}
    S_g=\displaystyle   \frac{\pi Q^{2} \left(4 \gamma + 1\right) \left(\gamma^{2} + 4\gamma+ 1\right)}{3 \gamma^{2}} \ln{\left(\frac{4 \left(\gamma + 1\right)}{4 \gamma + 1}\right)}.
\end{equation}
As stipulated before, the critical temperature $T_c$ and entropy $S_c$ are obviously obtained in the limit $\gamma\longrightarrow 1$, as
\begin{equation}\label{bhr21}
T_c= \frac{\sqrt{6}}{18 \pi Q},\quad \displaystyle S_c=10 \pi Q^{2}\ln{\left(\frac{8}{5} \right)}.
\end{equation}

The complete picture of the Maxwell construction in the $T_R-S_R$ diagram is depicted in right panel of  Fig.\ref{fig:mc_T_S}, 
As in the $P_R-V$ plane, the dashed black line is again the saturation line, and $S_{l,g}$ present the intersection of this curve with an isobar curve given for each value of $\gamma$. In the portion between these two values $S_l < S_R < S_g$, the black hole system is unstable. Thus, the oscillating part, below the critical temperature $T_c$, should be replaced with the isotherm $T_R=T_0$. On this isotherm $(T_R=T_0)$ the two black hole phases coexist. In a similar way, one notes that the intersection points meet and correspond to the critical horizon entropy/radius. Also, on the critical point $({\bm C})$, the two-phase system undergoes a second-order phase transition however a first-order phase transition occurs in all intersection points below $({\bm C})$. Once more we see that the supercritical phase dominates above $T_c$.\\

In the following, we complete the correspondence with the $VdW$ fluid by computing the latent heat of the charged black hole phase transition. Furthermore, a microscopic interpretation of such transition is attempted.

\section{Two-phase coexistence curves, latent heat, and the microscopic explanation of the phase change in Rényi formalism}\label{sect4}
\begin{figure}[!ht]
\centering
	\includegraphics[scale=.55]{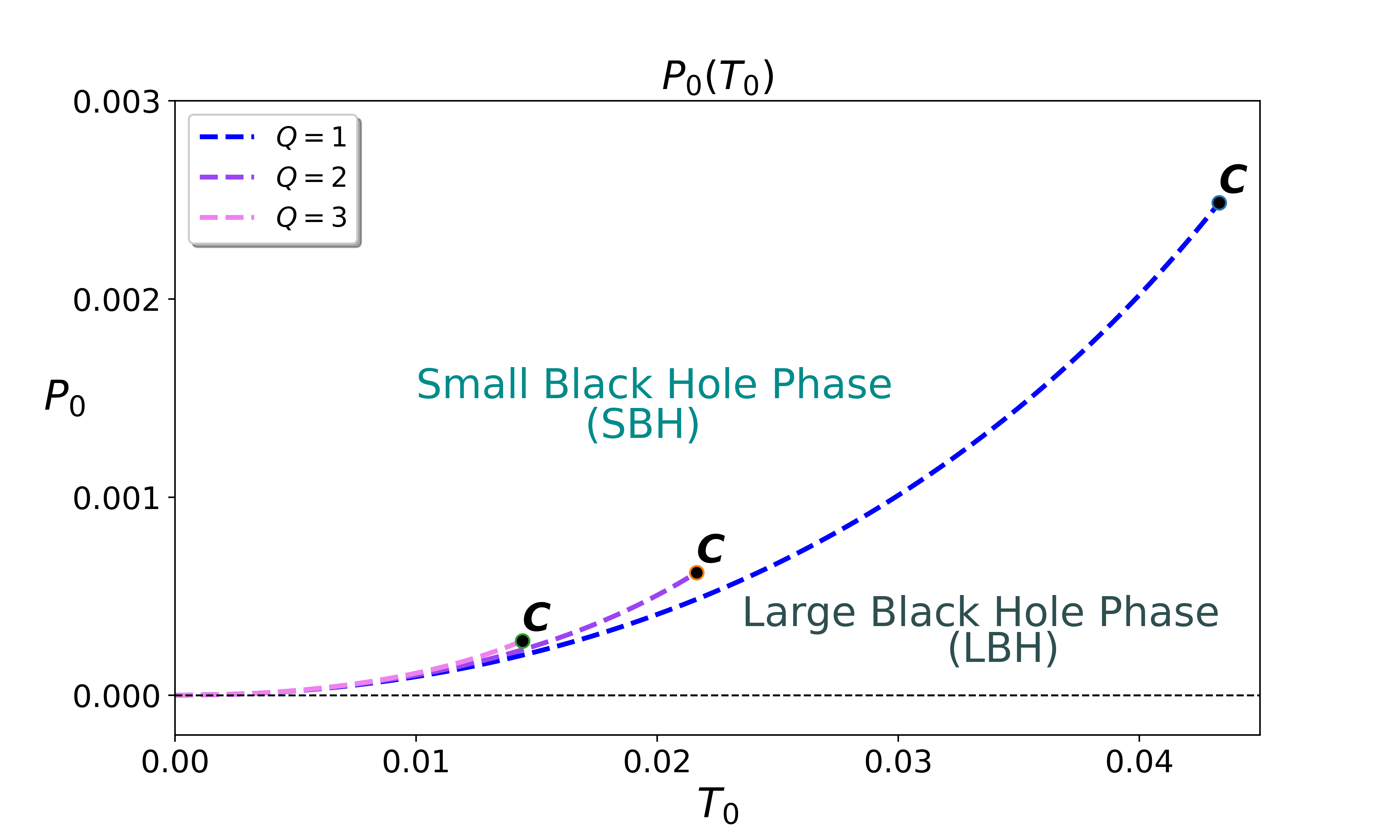}
	\vspace{-0.5cm}
	\caption{\footnotesize {\it $P_{0}-T_{0}$ coexistence curves for fixed charge $Q$.}}
	\label{fig:mc_co_curv}
\end{figure}

\paragraph{}It is well known that during a first-order phase transition of a $VdW$ fluid between its liquid and gas phases a latent heat is exchanged with the heat bath. The Clapeyron equation is the direct modelisation of these phase changes and within the correspondence examined in the previous sections between the $VdW$ fluid and the charged-flat black hole in Rényi formalism, we can write,
\begin{equation}\label{bhr22}
\frac{dP_0}{dT_0}=\frac{L_{l\longrightarrow g}}{T_0(v_g-v_l)},
\end{equation}
 where $L_{l\longrightarrow g}$ is the latent heat accompanying the phase transition from the black hole liquid-like phase to the gas-like phase (small/large). Herein, we attend to examine the two-phase equilibrium coexistence $P_0-T_0$ curves and the slope $\frac{dP_0}{dT_0}$ of those curves for the charged asymptotically flat black hole in the Rényi extended phase space. In Fig.\ref{fig:mc_co_curv} we illustrate the coexistence curves for different fixed charge values $Q$.
From such a figure, one can observe the effect of the electric charge $Q$ on the phase diagram, indeed each curve terminates by the critical point whose coordinates decrease by increasing the charge $Q$, while the slope $\frac{dP_0}{dT_0}$ becomes more important as the electric charge grows. Moreover, the Fig.\ref{fig:mc_co_curv} reveals also that
the pressure $P_0$ tends toward zero with decreasing temperature $T_0$ and that the electric charge has small effect on the coexistence curves for small values of temperature and pressure. \\

\begin{figure}[!ht]
		\centering
			\begin{tabbing}
			\centering
			\hspace{-1.5cm}
			\includegraphics[scale=.4]{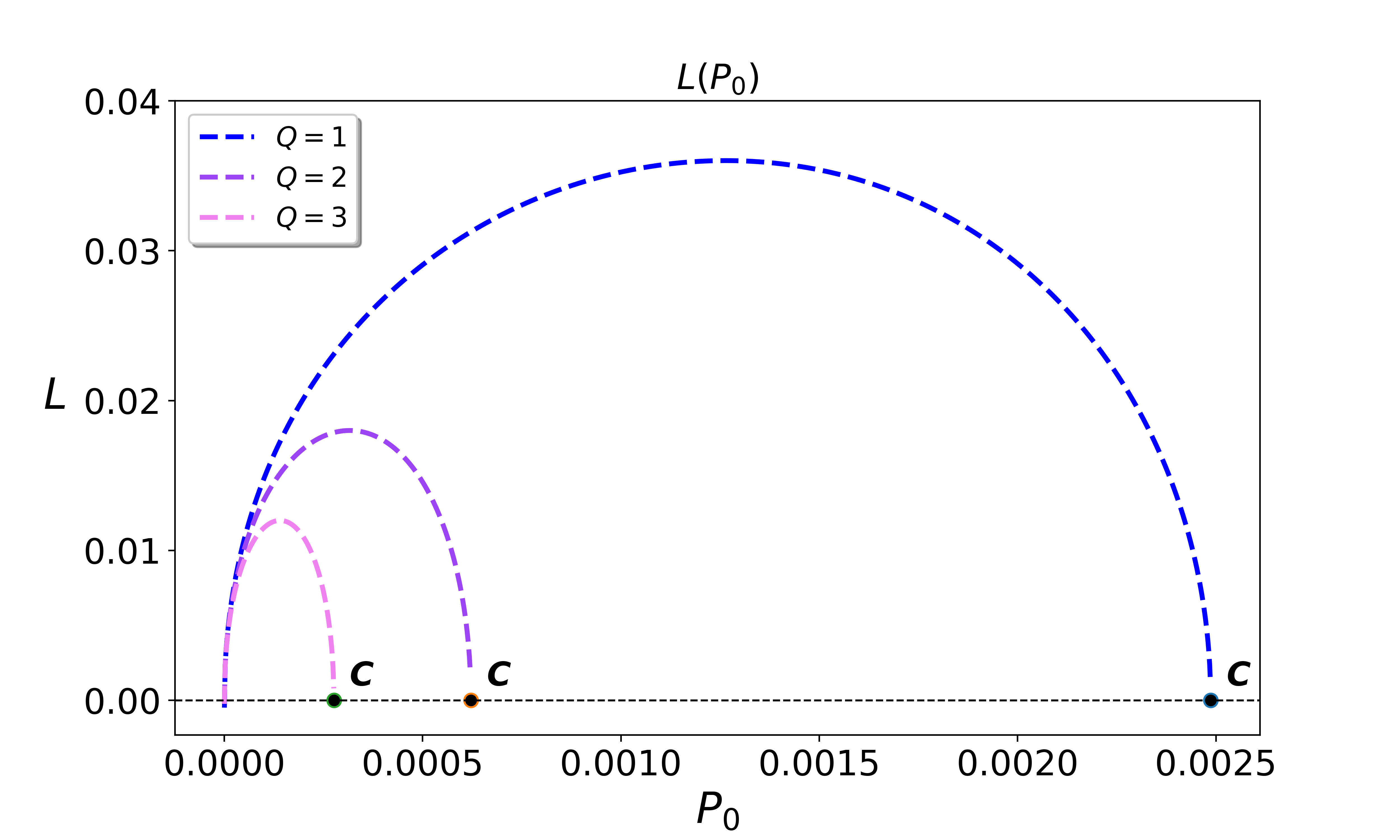} 
			\hspace{-1.2cm}
			\includegraphics[scale=.4]{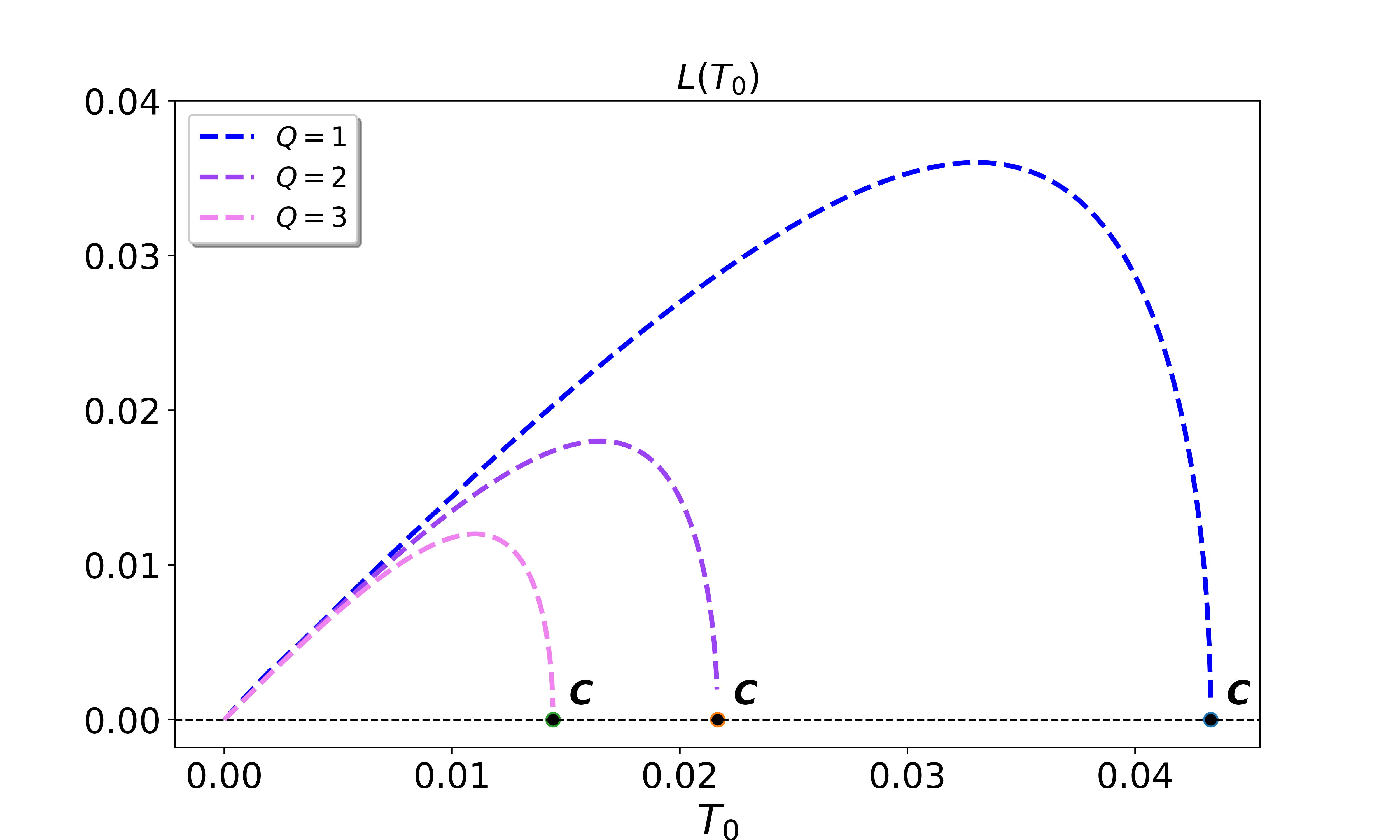} 
		   \end{tabbing}
		   \vspace{-1.cm}
 	      \caption{\footnotesize{\it The variation of the latent heat $L$ with the coexistence pressure $P_0$ and temperature $T_0$ for different values of charge $Q$.} }
	\label{fig:mc_L}
\end{figure}

\begin{figure}[!ht]
    \centering
	\includegraphics[scale=.5]{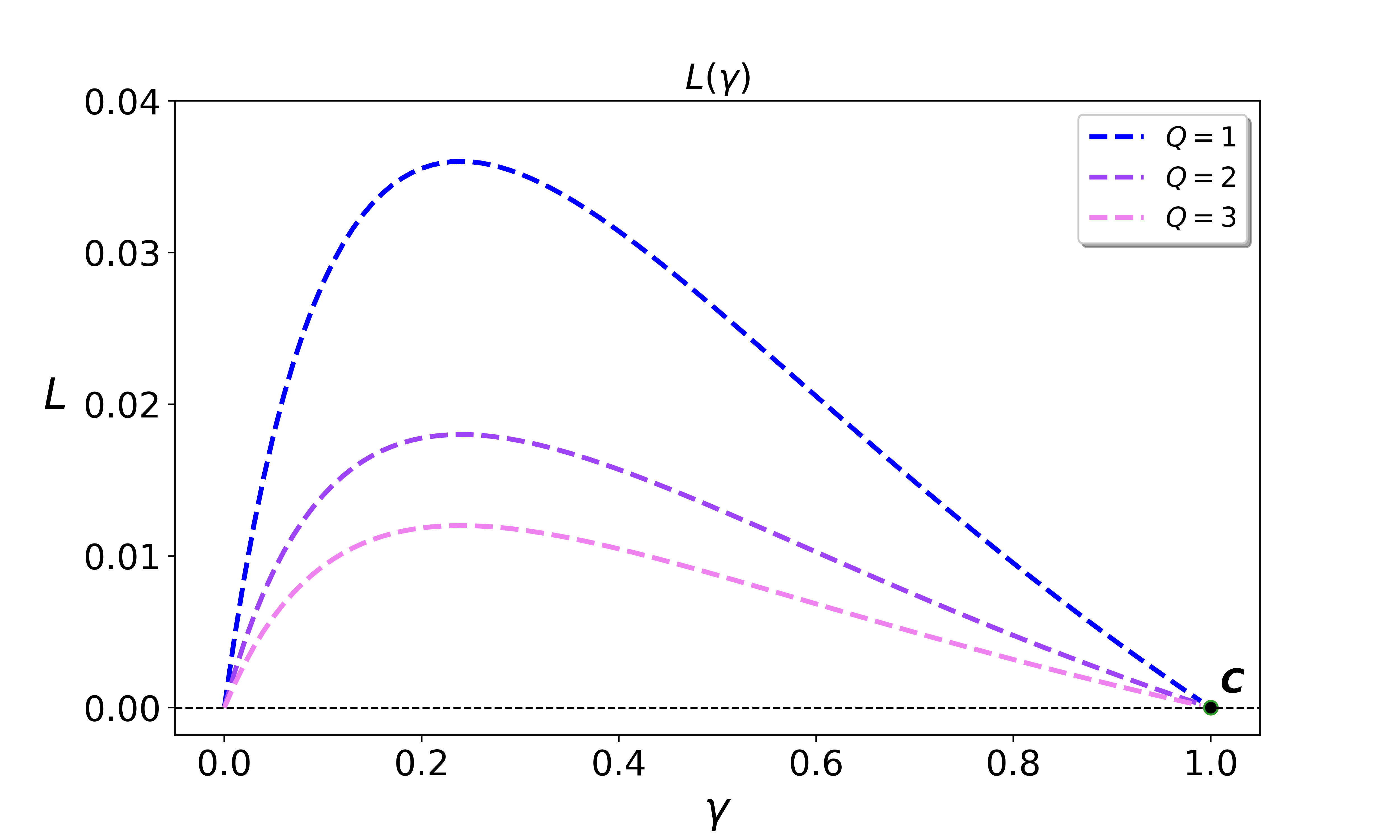}
	\caption{\footnotesize{\it The variation of the latent heat $L$ in terms of the critical ration $\gamma$  for different fixed  values of charge $Q$.}}
	\label{fig:mc_L_gamma}
\end{figure}

Solving Eq.\eqref{bhr22} for the latent heat and using Eq.\eqref{bhr12} we obtain the latent heat of the phase transition of RN-flat black hole in Rényi formalism as,
\begin{equation}\label{bhr23}
L(\gamma)= \frac{3\gamma (1-\gamma )   (\gamma +1)^2 }{2 \pi  Q\left(\gamma ^2+\gamma +1\right) (\gamma^2+4\gamma+1)^{3/2}}.
\end{equation}\\
The variation of the latent heat with the pressure $P_0$ and the temperature $T_0$ is depicted in Fig.\ref{fig:mc_L} for various values of charges, while its behaviour in terms of the ratio $\gamma$ is illustrated in Fig.\ref{fig:mc_L_gamma}.

For both figures, one can observe the effect of the pressure $P_0$, the temperature $T_0$, and the ratio $\gamma$  on the latent heat of phase transition. In fact, when all the previous quantities grow the latent heat $L$ is not a monotonic function but increases firstly and then decreases to reach zero at the critical point ($P_R\to P_c$/ $T_R\to T_c$/ $\gamma\to 1$), where a second phase transition takes place. Furthermore, the latent heat $L$ decreases the increasing the charge.



Nowadays, a variety of papers in the literature are devoted to the microstructure of the phase transition of black holes where one links the different sizes of a black hole to a different density of its molecules \cite{Wei:2015iwa,Wei:2019uqg,Guo:2019hxa,Chabab:2017xdw,Wei:2019yvs,Miao:2017fqg}.  
Otherwise, the Landau theory of continuous phase transition is typified by changes in the degree of the material order and the accompanying changes in the symmetry of matter. Thus the analogy between the black hole thermodynamics and the ordinary one pushes us to think about the symmetry change in the black hole and the phase transition within the Rényi formalism. In what follows, we put such assertion under investigation by considering a charged-flat black hole solution. 

\begin{figure}[!ht]
		\centering
			\begin{tabbing}
			\centering
		\hspace{-1.4cm}
			\includegraphics[scale=.4]{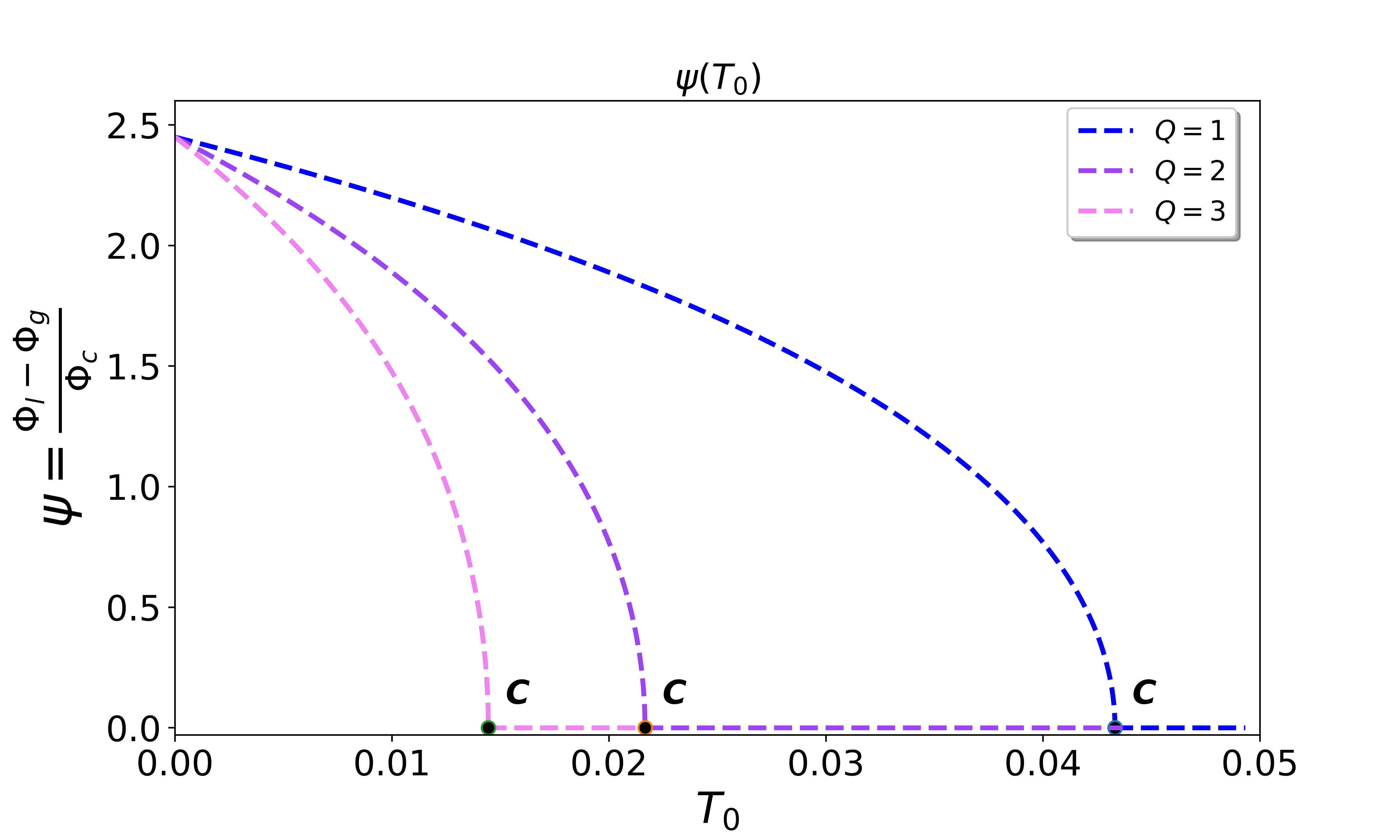}
			\hspace{-1.6cm}
			\includegraphics[scale=.4]{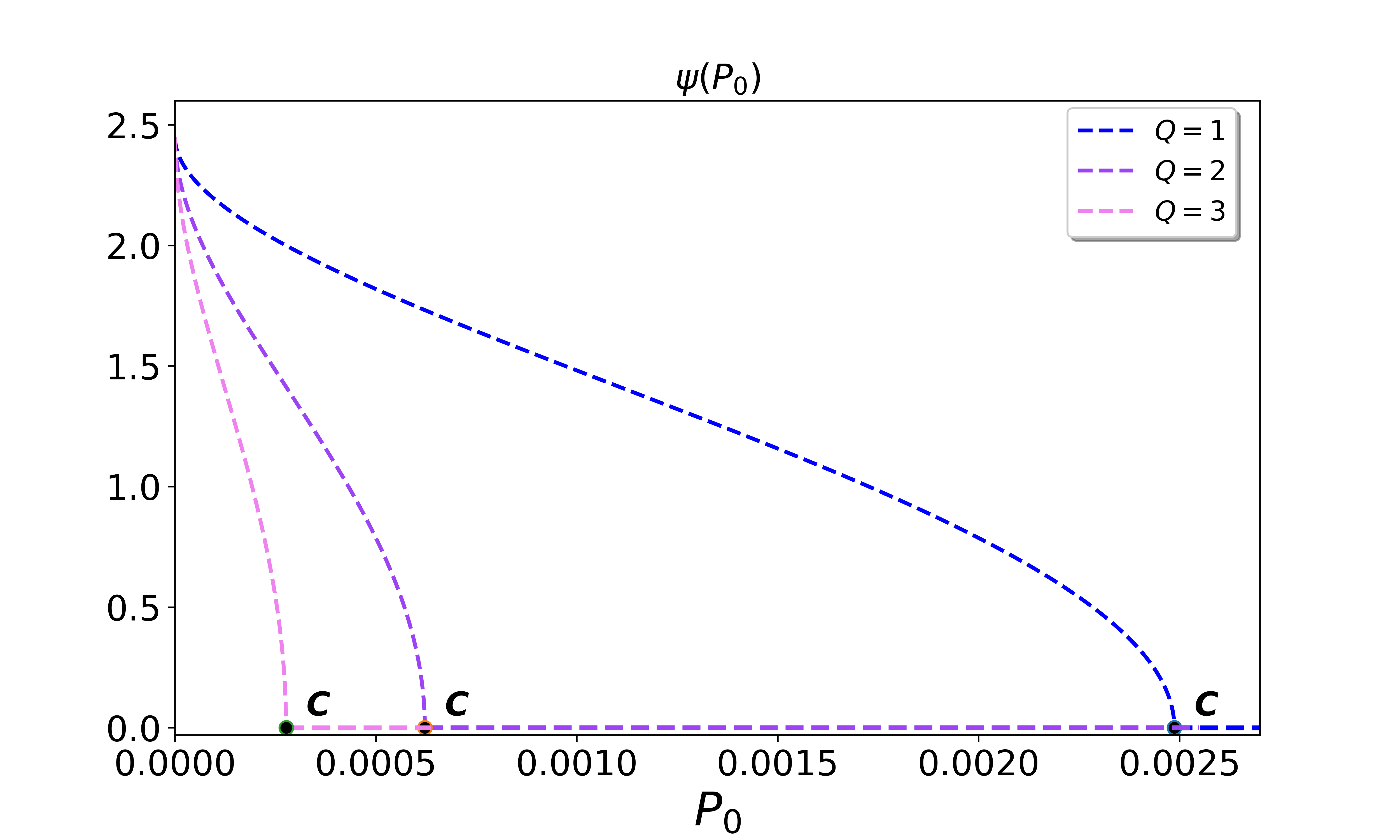}
		   \end{tabbing}
		   \vspace{-1.cm}
 	      \caption{\footnotesize{\it The behaviour of the order parameter $\psi$ in terms of the coexistence temperature $T_0$ (left panel) and the coexistence pressure $P_0$ (right panel) for different values of charge $Q$.} }
	\label{fig:mc_psi}
\end{figure}

During the black hole phase transition, the potential $\Phi=\frac{Q}{r_h}$ presents a mutation, which unveils the conflicting microstructures of the black hole molecules in different phases. 
The electric potentials of the two-phase system are, respectively given by
\begin{equation}\label{bhr24}
    \Phi_l=\frac{Q}{r_l}, \quad \Phi_g=\frac{Q}{r_g}.
\end{equation}
When $T_0\leq T_c$ or $P_0\leq P_c$ which means  $0<\gamma\leq1$, one defines the order parameter as:
\begin{equation}\label{bhr25}
\Psi(T_0)=\frac{\Phi_l-\Phi_g}{\Phi_c}= \frac{\sqrt{6} \left(1 - \gamma\right)}{\sqrt{\gamma^{2} + 4 \gamma + 1}}.
\end{equation}\\
Fig.\ref{fig:mc_psi} shows a characteristic behavior of the order parameter $\Psi$ as a function of the coexistence temperature $T_0$ and pressure $P_0$ for a critical exponent $\beta=\frac{1}{2}$. A Taylor series expansion in the vicinity of $T_c$ and $P_c$,  Eqs.\eqref{bhr26}, confirms the value of $\beta$.

\begin{equation}\label{bhr26}
\left\{\begin{array}{c}\Psi(T_0)=  2 \cdot 6^{\frac{3}{4}} \sqrt{\pi} \sqrt{Q} \sqrt{ T_c- T_{0}}  + \mathcal{O}\left(T_{0} - T_c\right) \quad (T_0<T_c)\\
\Psi(P_0)=\displaystyle  8 \sqrt{6} \sqrt{\pi} Q \sqrt{P_c- P_{0}}  + \mathcal{O}\left(P_{0}-P_c\right)\quad (P_0<P_c)
\end{array}\right.
\end{equation}

When the black hole exhibits a temperature or pressure under the critical ones and the black hole molecule is at a high potential phase $1$, molecules undergoing such a potential $\Phi$ align in a certain orientation. 
In the low symmetry case, the black hole molecules are quite ordered, when the black hole switches to phase $2$ within the same temperature/pressure, the potential $\Phi$  reduces and consequently, the molecule orientation gets confused leading to higher symmetry.
Moreover,  the phase below the critical temperature, possesses low symmetry, higher order, and non-zero $\Psi$, whereas, the phase above the critical temperature shows higher symmetry, lower order, and the order parameter $\Psi$ is zero.

According to Landau's perspective, the order parameter $\Psi$ is a small amount near the critical point $T_c$ and the Gibbs energy $G(T_R,\Phi)$ can be expanded as a power  $\Psi$ as \cite{Guo:2019hxa},
\begin{equation}
\label{eq35}
G(T_R,\Phi ) = G_0 (T_R) + \frac{1}{2}a(T_R)\Psi ^2 + \frac{1}{4}b(T_R)\Psi ^4 +
\cdots ,
\end{equation}
in which $G_0(T_R)$ stands for the Gibbs function at $\Psi=0$. The reason behind the presence of only even order terms of the order parameter is that the  system is invariant under the parity transformation $\Psi\rightleftarrows -\Psi$. The Gibbs function presents three minimal values located at
\begin{equation}
\label{eq36}
\Psi = 0,
\quad
\Psi = \pm \sqrt { - \frac{a}{b}}.
\end{equation}
The first trivial solution $\Psi=0$ is associated with the disordered state, corresponding to $T_R>T_c$ temperature range at $a>0$, while the  non vanishing solutions $\Psi=\pm\sqrt{-\frac{a}{b}}$ correspond to an orderly state, when $T_R<T_c$ at $a<0$. Although we have $a=0$ at $T_R=T_c$, near the critical point one can write
\begin{equation}
\label{eq37}
a = a_0 \left( {\frac{T_R - T_c }{T_c }} \right) = a_0 t.
\quad
a_0 > 0,
\end{equation}
since $\Psi=\pm \sqrt{-\frac{a}{b}}$ is a real quantity. When $T_R<T_c$, $a<0$,
so  $b$ should be  positive and we have
\[
\Psi = 0,
\quad
t > 0,
\]
\begin{equation}
\label{eq38}
\Psi = \pm \left( {\frac{a_0 }{b}} \right)^{1 / 2}( - t)^{1 / 2},
\quad
t < 0.
\end{equation}
Several ferromagnetic systems share the following experimental features near the critical point field, in fact,  

\begin{itemize}
\item At $t \to 0^- $, the spontaneous magnetization behaves as
\begin{equation}
\label{eq39}
\mathcal{M} \propto ( - t)^\beta ,
\quad
t \to 0^- .
\end{equation}

\item  
The zero field magnetic susceptibility of various ferromagnetic substances $\chi = \left( {\frac{\partial \mathcal{M}}{\partial H}}
\right)_{T_R} $ presents a singularity at $t \to 0^{\pm}$  and in the vicinity of these singularities, $\chi$ varies as,
\begin{equation}
\label{eq40}
\chi \propto t^{ - \zeta },
\quad
t \to 0^{+};
\quad
\chi \propto ( - t)^{ - \zeta '},
\quad
t \to 0^{-}.
\end{equation}

\item While, at $t=0$ and very weak magnetic field, the magnetization $\mathcal{M}$ is linked to  the external magnetic field  $H$ by the following law
\begin{equation}
\label{eq41}
\mathcal{M} \propto H^{1 / \delta }.
\end{equation}

\item Additionally, when $t\to 0^\pm$, the zero field specific heat capacity of ferromagnetic material $c_H(H = 0)$ obeys the following behavior
\begin{equation}
\label{eq42}
c_H \propto t^{ - \bar {\alpha }},
\quad
t > 0;
\quad
c_H \propto ( - t)^{ - \bar {\alpha }'},
\quad
t < 0.
\end{equation}
\end{itemize}
The relation $\Psi(t)$ is the same between  Eq.(\ref{eq38}) 	and  Eq.(\ref{eq39}), and give the critical index as $\beta=1/2$. Following  the method
of Guo .et al~\cite{Guo:2019oad,Guo:2019hxa}, we derive the critical exponent as $\bar{\alpha}=\bar{\alpha }'=0$, $\zeta=\zeta'=1$, $\delta=3$,
and the Rényi entropy of RN-flat black hole near the critical point.\\

The Rényi entropy of disordered phase is $S_{R_{diso}}$, and the Rényi entropy of ordered phase is
\begin{equation}
\label{eq43}
S_{R_{ord}}=S_{R_{diso}}+\frac{a_0^2 t}{2bT_c}.
\end{equation}
This equation traduces the continuity of the Rényi entropy at the critical point. Concretely
\begin{equation}
S_{R_{ord}}=S_{R_{diso}}, \qquad \text{ at } t=0.
\end{equation}
This Landau's formalism behavior is in the same way as that of the charged black hole in anti-de Sitter space with the standard Gibbs-Boltzmann statistics.\\

In the next section, we introduce a new mathematical approach aiming to put on more firm ground the construction of the extended phase space of the charged-flat black hole in the Rényi formalism.

\section{Hamiltonian approach to Rényi's thermodynamics of charged-flat black hole}\label{sect5}

 	
 \paragraph{} Gauge/Gravity duality posits a correspondence between gauge and gravity theories.  As pointed out in the introduction, given that the gauge theory side of the duality admits a standard thermodynamic treatment, it is natural to expect that the gravity side too supports a standard thermodynamics approach. Such a picture can be fulfilled 
 by increasing the number of degrees of freedom of black holes, thus, leading to a thermodynamics in accordance with standard thermodynamics of matter systems. 
 The Hamiltonian approach\cite{Baldiotti:2016lcf, Baldiotti:2017ywq,Ghosh:2019rsu} to thermodynamics, which we attend to apply in this section, is a new and powerful scheme to investigate and extend the phase space of black holes.

\paragraph{}In the Hamiltonian approach to thermodynamics, one considers all equations of state of a given thermodynamic system as constraints on phase space \cite{Baldiotti:2016lcf,Baldiotti:2017ywq}. For each thermodynamic potential $M$, its differential $dM$ is expressed by the canonical and tautological form $p dq$ on the constraint surface defined by these equations of state. Then it is possible to extend the phase space by the introduction of a canonical conjugate pair $(\theta, \tau )$ such that the form $pdq + \theta d\tau$ reduces to the \textit{Poincaré-Cartan} form $p dq - h d\tau$ on the constraint surface $H=\theta+h(q,p,\tau)$. Thus, one obtains a description in both spaces, the reduced phase space $(p, q)$  and the extended phase space $(p, q; \theta, \tau )$. Therefore, all thermodynamic potentials are related by canonical transformations, giving equivalent representations. In this way, one is able to increase the degrees of freedom of the thermodynamic system.
 Further, through the general Hamiltonian approach to RN-AdS black hole, we can verify that the thermodynamical volume conjugate to the thermodynamical pressure is indeed equal to the volume of a sphere of radius $r_h$, such that $PV=-V\frac{\Lambda}{8\pi}=-r_h^3\frac{\Lambda}{6}$ is interpreted as the energy extracted from spacetime due to the presence of the black hole\cite{Johnson:2014yja}, such calculations are based on the promotion of the cosmological constant $\Lambda$ to a function of coordinates in the phase space through a new equation of state. Now, it's legitimate to check  whether the volume of the RN-flat black hole  persists within Rényi statistics and by considering that the nonextensivity parameter $\lambda$ undergoes the same assumptions as $\Lambda$.

\paragraph{}According to the first law of black hole thermodynamics, the minimal RN-AdS in Gibbs-Boltzmann has the entropy as the only free thermodynamical variable, which makes it a one-dimensional system. Therefore, any minimal mechanical analog should be one-dimensional as well. A direct identification between mechanical $(p,q)$, and thermodynamic variables $(T_H,S_{BH})$, up to canonical transformations, is \cite{Haldar:2019pwt}
\begin{equation}\label{analog_mec}
q=\frac{S_{BH}}{\pi},\quad p=\pi T_H,
\end{equation}
 where $S_{BH}$ and $T_H$ are the Bekenstein-Hawking entropy and the Hawking temperature of a RN-AdS black hole respectively given as,
\begin{equation}\label{eq_state_AdS}
T_H=\frac{1}{4\pi r_h}-\frac{Q^2}{4\pi r_h^3}-\frac{\Lambda r_h}{4\pi} \quad \text{and} \quad
S_{BH}=\pi r_h^2.
\end{equation}
\paragraph{}In Rényi formalism, a similar identification can be made. Indeed, 
 recalling the expressions of the Rényi entropy $S_R$, Eq.\eqref{bh17} and of the Rényi temperature $T_R$ Eq.\eqref{Tr}, we can define the Rényi mechanical analog of RN-flat black hole in Rényi statistics by the identification of $S_R$ and $T_R$ with  $q_{\lambda}$ and $p_{\lambda}$ respectively as,
\begin{equation}
   q_{\lambda}=\frac{S_R}{\pi}=q-\frac{\pi \lambda q^2 }{2},\quad p_{\lambda}=\pi T_R=p(1+\pi\lambda q).\label{transformation}
\end{equation}
In which, we have expressed the Rényi mechanical variables in terms of the mechanical Gibbs-Boltzmann $(GB)$ variables to a first order in the nonextensivity parameter $\lambda$. In $GB$ formalism, the differential of the RN-AdS black hole mass $M=\frac{r_h}{2}(1+\frac{Q^2}{r_h^2}-\frac{\Lambda r_h^2}{3})$, taken here as the thermodynamical potential, in terms of the mechanical variables Eqs.\eqref{analog_mec}, reads for fixed charge $Q$,\\
\begin{equation}\label{dM}
     dM=pdq-\frac{1}{6}q^{\frac{3}{2}}d\Lambda.
\end{equation}
where $\Lambda$ is the cosmological constant considered in the Hamiltonian approach as a function a priori of all mechanical variables. The one-form Eq.\eqref{dM} is constrained by the equation of state of a RN-AdS black hole, Eq.\eqref{eq_state_AdS} written in mechanical variables as,
\begin{equation}\label{constraint_AdS}
     4p=\frac{1}{\sqrt{q}}-\frac{Q^2}{q^{\frac{3}{2}}}-\Lambda \sqrt{q}
\end{equation}

The conjectured equivalence between the RN-AdS black hole in Gibbs-Boltzmann formalism and RN-flat black hole in Rényi formalism\cite{Czinner:2015eyk, Czinner:2017tjq} permits one to write the differential of the RN-flat black hole mass in Rényi statistics in terms of the Rényi mechanical variables defined by Eq.\eqref{transformation}, for fixed charge $Q$ as,
\begin{equation}\label{dM_renyi}
  dM=T_RdS_R=p_{\lambda}dq_{\lambda}.
\end{equation}
Also from Eq.\eqref{transformation} we compute the differential of $q_{\lambda}$, we get,
\begin{equation}\label{dq_lam}
dq_{\lambda}=(1-\pi\lambda q)dq-\frac{\pi q^2}{2}d\lambda.
\end{equation}
where $\lambda$, similarly to $\Lambda$, is promoted to a function of the mechanical variables.
Next, substituting Eq.\eqref{dq_lam} and \eqref{transformation} in Eq.\eqref{dM_renyi} and keeping leading order terms in $\lambda$, we obtain for $dM$,
\begin{equation}\label{one_form_rényi}
dM=pdq-p(1+\pi\lambda q)\frac{\pi q^2}{2}d\lambda.
\end{equation}
 Eq.\eqref{one_form_rényi} is the one-form of the thermodynamical potential $M$, in the Rényi formalism.
 By the help  of the state's equation in Rényi formalism Eq.\eqref{bh25} which is now interpreted as a constraint on the phase space $(q,p)$ and  applying the transformation equations Eqs.\eqref{transformation}, the equation of state reduces to,
\begin{equation}\label{constraint_rényi}
4p=\frac{1}{\sqrt{q}}-\frac{Q^2}{q^{\frac{3}{2}}}.
\end{equation}

Thus, in Rényi formalism and for a RN-flat black hole in four spacetime dimensions, the Hamiltonian approach is based on the one-form Eq.\eqref{one_form_rényi} subjected to the constraint Eq.\eqref{constraint_rényi}. In what follows,  
we will apply such an approach to  RN-flat black hole within the Rényi statistics.
\paragraph{}The starting point is to promote the nonextensivity parameter $\lambda$ to a function on phase space of the coordinate $q$, $\lambda=\lambda(q)$. The one-form $dM$, Eq.\eqref{one_form_rényi} becomes,

\begin{equation}
    dM=\left[p-p(1+\pi\lambda q)\frac{\pi q^2}{2}\frac{\partial\lambda}{\partial q}\right]dq.
\end{equation}
The expression of $dM$ is in the form $\alpha_{\lambda}=\omega_{\lambda} dq$, where $\omega_{\lambda}$ is obtained to be:
\begin{equation}\label{T_eff_rényi}
\omega_{\lambda}=p-p(1+\pi\lambda q)\frac{\pi q^2}{2}\frac{\partial\lambda}{\partial q},
\end{equation}
and restricted to the constraint surface in phase space given by the equation of state Eq.\eqref{constraint_rényi} as:
\begin{equation}\label{constraint_phi}
    \phi_{\lambda}=4p-\frac{1}{\sqrt{q}}+\frac{Q^2}{q^{\frac{3}{2}}}=0.
\end{equation}
Which allow us to write the compact notation,
\begin{equation}
dM=\alpha_{\lambda}|_{\phi_{\lambda}=0}
\end{equation}
and define the symplectic $2$-form
\begin{equation}
\Omega_{\lambda}=\left(\frac{\partial\omega_{\lambda}}{\partial p}\right)dq\wedge dp,
\end{equation}
to make the transformation $(p,q)\rightarrow(\omega_{\lambda},q)$ canonical. \textit{The extension of phase space is accomplished by the introduction in $dM$ of a new pair of conjugate variables say, $(\theta_{\lambda},\tau_{\lambda})$ such that,}
\begin{equation}\label{extension_rényi}
dM=\omega_{\lambda} dq+\theta_{\lambda} d\tau_{\lambda}.
\end{equation}
Taking this time $\lambda$ as a function of $q$ and $\tau_{\lambda}$, $\lambda=\lambda(q,\tau_{\lambda})$, one can write,
\begin{equation}\label{subtitution_rényi}
d\lambda=\frac{\partial\lambda}{\partial q}dq+\frac{\partial\lambda}{\partial \tau_{\lambda}}d\tau_{\lambda}.
\end{equation}
By the help of equation Eq.\eqref{subtitution_rényi}, Eq.\eqref{one_form_rényi} is re-expressed as
\begin{equation}\label{extension2_rényi}
dM=\omega_{\lambda}dq-p(1+\pi\lambda q)\frac{\pi q^2}{2}\frac{\partial\lambda}{\partial \tau_{\lambda}}d\tau_{\lambda}.
\end{equation}
Confronting  Eq.\eqref{extension_rényi} to Eq.\eqref{extension2_rényi} leads to the addition of a new constraint on the extended phase space as,
\begin{equation}\label{constraint_add_rényi}
H_{\lambda}=\theta_{\lambda}+p(1+\pi\lambda q)\frac{\pi q^2}{2}\frac{\partial\lambda}{\partial \tau_{\lambda}}=0.
\end{equation}

The constraint $H_{\lambda}=0$ reduces the $1$-form Eq.\eqref{extension_rényi} in the extended phase space to the \textit{Poincaré-Cartan} form $p dq-h_{\lambda} d\tau_{\lambda}$ in reduced phase space such that,
\begin{equation}
h_{\lambda}(q,\tau_{\lambda})=p(1+\pi\lambda q)\frac{\pi q^2}{2}\frac{\partial\lambda}{\partial \tau_{\lambda}}.
\end{equation}
We define a symplectic 2-form in the extended phase space which preserves the canonical relations among the transformed coordinates $X_{\lambda}=(\omega_{\lambda},q;\theta_{\lambda},\tau_{\lambda})$ such as,
\begin{equation}
\Bar{\Omega}_{\lambda}=\left(\frac{\partial\omega_{\lambda}}{\partial p}\right)dq\wedge dp+\left(\frac{\partial\omega_{\lambda}}{\partial \tau_{\lambda}}\right)dq \wedge d\tau_{\lambda}+d\tau_{\lambda}\wedge d\theta_{\lambda}.
\end{equation}

Now, we are able to generate  the canonical \textit{Poisson brackets}, for a pair of functions of the coordinates say, $f_{\lambda}(X_{\lambda})$ and $g_{\lambda}(X_{\lambda})$ we have, 
\begin{equation}
\{f_{\lambda},g_{\lambda}\}=\Bar{\Omega}_{\lambda}(\xi_{f_{\lambda}},\xi_{g_{\lambda}}),
\end{equation}
where $\xi_{f_{\lambda}}$ and $\xi_{g_{\lambda}}$ are vector fields generated from $f_{\lambda}$ and $g_{\lambda}$ respectively.\\

The identification of $\tau_{\lambda}$ with Rényi thermodynamical pressure $P_R$, leads to $dM=T_{eff,R}\:dS_R$ for constant $P_R$, where $T_{eff,R}$ is the thermodynamical temperature given by Eq.\eqref{T_eff_rényi} such as,
\begin{equation}
    T_{eff,R}=\displaystyle\frac{\omega_{\lambda}}{\pi}. 
\end{equation}

Thus $dM$ is the exchanged heat in an isobaric transformation. Therefore we identify $M$ with the enthalpy of the black hole. From the constraint Eq.\eqref{constraint_add_rényi}, we read the thermodynamical volume $V$ conjugate  to $P_R$ in Rényi thermodynamics as
\begin{equation}\label{volume_rényi_q}
V=p(1+\pi\lambda q)\frac{\pi q^2}{2}\frac{\partial\lambda}{\partial P_R}.
\end{equation}
In the Rényi extended phase space for a RN-flat black hole in the canonical ensemble, the parameter $\lambda$ is associated with the thermodynamic Rényi pressure $P_R$ and the electric charge $Q$ as \cite{Promsiri:2020jga}
\begin{equation}\label{P_rényi_lam_Q}
    P_R=\frac{3\lambda}{32}(1-\frac{Q^2}{q})\implies \lambda=\frac{32 P_R}{3(1-\frac{Q^2}{q})}.
\end{equation}
A substitution of Eq.\eqref{P_rényi_lam_Q} in Eq.\eqref{volume_rényi_q} and using the constraint Eq.\eqref{constraint_phi} and the definition $q=r_h^2$ in Eq.\eqref{analog_mec} gives
\begin{equation}\label{volume_rényi}
V=\frac{4\pi}{3}r_h^3+\mathcal{O}(\lambda).
\end{equation}
Since the Rényi pressure $P_R$ is proportional to $\lambda$, the correction term to the thermodynamical volume of order $\mathcal{O}(\lambda)$ would generate a second order correction $\mathcal{O}(\lambda^2)$ to the expression of the enthalpy $M$ Eq.\eqref{Smarr_rényi_mod}, which can be neglected\cite{Promsiri:2020jga} owing to the condition $0<\lambda<<1$. The energy extracted from spacetime due to the presence of the black hole is given by
\begin{equation}
  P_RV= \frac{\pi\lambda}{8}r_h^3(1-\frac{Q^2}{r_h^2}).
\end{equation}
In the large $r_h$ limit $(r_h>>Q)$ , the functional dependence on $r_h$ of this energy matches the RN-AdS case, $\displaystyle-\frac{\Lambda}{6}r_h^3$,  with the identification $|\Lambda|\rightarrow\frac{3\pi}{4}\lambda$.

\paragraph{}The application of the Hamiltonian approach to the RN-AdS black hole in Gibbs-Boltzmann formalism and to RN-flat black hole in Rényi formalism presents a great similarity and strengthens once more the conjectured equivalence between these two systems with the advantages of the Rényi statistics outlined in the introduction of the present paper.

\section{Conclusion}\label{sect6}
\paragraph{}In this paper, we have investigated some phase equilibrium features of charged black holes in flat spacetime via Rényi statistics. We first reviewed the thermodynamical structure in such a background. This phase portrait is similar to the Van-der-Waals one and the charged-AdS black holes phase picture in the Gibbs-Boltzmann statistics. Concretely, we have shown that the oscillatory behaviour persists in the $P_R- V$ and $T_R-S_R$ diagrams. Afterwards, by means of Maxwell's equal-area law, the unphysical branch of the system has been excluded and the phase transition point has been disclosed.
Furthermore, we have established the Clapeyron equation at the coexistence curve associated with each diagram, which puts light on the latent heat of the phase change and its behaviour under varying the black hole charge $Q$.

A pertinent comprehension of the phase structure of charged-flat black hole thermodynamics via Rényi statistics can help to learn about the similitude with the charged-AdS black hole in Gibbs-Boltzman formalism and a possible profound connection between the non-extensive parameter $\lambda$ and the cosmological constant $\Lambda$. Putting such assertion under more deep investigations, Landau's continuous phase transition theory has been used to inspect the critical behaviour of such a black hole in the Rényi's formalism, the critical index has been also obtained. Meanwhile, both concepts confirm such similitude. \\

Lastly, we have reconstructed the previous thermodynamics results by the Hamiltonian approach promoted to the Rényi statistics framework.  Concretely,  we have taken the nonextensivity parameter $\lambda$ as a function of coordinates of the phase space via a new thermodynamical equation of state based only on the homogeneity of thermodynamics variables, such an equation of state exhibits the generalized thermodynamic volume. All these results consolidate the expected and possible bridge between the nonextensivity Rényi parameter $\lambda$ and the cosmological constant $\Lambda$ suggested for the first time in \cite{Promsiri:2020jga} and reinforced in  \cite{dilaton,Hirunsirisawat:2022ovg}.\\

 This study opens up further research perspectives.
It would be fascinating to make contact with Euclidean action formalism. Concretely, it is well known that the Rényi extension of standard black hole thermodynamics can be seen as generating conical defects in the Euclidean time coordinate \cite{Banados:1993qp}. This fact sparked significant research that broadens the notion of gravitational entropy  \cite{Faulkner:2013ana} and, in particular, offers a general method for calculating Holographic Entanglement Entropy. In addition, It would be intriguing to investigate the current Rényi phase transitions from the holographic point of view \cite{Nguyen:2015wfa,ElMoumni:2018fml}.
On all of the open issues mentioned, we plan to report in future works.

\bibliographystyle{unsrt}
\bibliography{maxwellR.bib}

\end{document}